\documentstyle[makeidx,fleqn,11pt,epsfig]{article}
\input{amssym.def}
\input{amssym.tex}
\author{ }
\date{ }
\makeatletter

\def\gsim{\compoundrel>\over\sim}
\def\lsim{\compoundrel<\over\sim}
\def\compoundrel#1\over#2{\mathpalette\compoundreL{{#1}\over{#2}}}
\def\compoundreL#1#2{\compoundREL#1#2}
\def\compoundREL#1#2\over#3{\mathrel
      {\vcenter{\hbox{$\m@th\buildrel{#1#2}\over{#1#3}$}}}}
\begin{document}
%
\vspace*{0.5cm}
\hfill\ {\Large\bf KEK Preprint 2008-33}

\vspace*{2mm}
\hfill\ {\Large\bf October 2008~~~~~~~~~~~~}

\vspace*{2mm}
\hfill\ {\Large\bf A/H\,~~~~~~~~~~~~~~~~~~~~~~~}

\vspace*{3.0cm}
\begin{center}
{\Huge A~~160$-$320\,GeV~~linear~~collider}\\ 
\vspace*{0.8cm}
{\Huge to~~study}\\
\vspace*{0.6cm}
{\Huge $e^{+}e^{-}\rightarrow$~HZ~~and~~$\gamma\gamma\rightarrow$~H,\,HH}\\
\vspace*{3.4cm}
{\Large R. BELUSEVIC}\\
\vspace*{0.8cm}
{\large IPNS, {\em High Energy Accelerator Research Organization} (KEK)}\\
\vspace*{1.3mm}
{\large 1-1 {\em Oho, Tsukuba, Ibaraki} 305-0801, {\em Japan}} \\
\vspace*{1.3mm}
{\large belusev@post.kek.jp}\\
\end{center}

\thispagestyle{empty}

\newpage

\vspace*{8cm}
\begin{center}
\begin{minipage}[t]{11.5cm}
``{\Large\em The changing of bodies into light, and light into bodies, is very
conformable to the course of nature, which seems delighted with
transmutations.}" 

\addvspace{5mm}
\hfill\ {\Large Isaac Newton}
\end{minipage}
\end{center}

\newpage

\tableofcontents
\addtocontents{toc}{\protect\vspace{1.3cm}}
\vspace*{5mm}
\noindent
{\large\bf References}


\newpage

\begin{center}
\begin{minipage}[t]{13.6cm}
{\bf Abstract\,:}\hspace*{3mm}
{The construction of two electron linacs and an optical FEL system is
proposed. This facility, which would serve primarily as a Higgs-boson collider 
factory, could be built in two stages, each with distinct physics objectives
requiring particular center-of-mass (CM) energies: 
(1) $e^{+}e^{-}\rightarrow{\rm HZ}$ (${\rm E}_{e^{+}e^{-}}\sim 250~{\rm GeV})$,
and (2) $\gamma\gamma\rightarrow{\rm H,\,HH}$ (${\rm E}_{e^{-}e^{-}} \sim
160~{\rm to}~320~{\rm GeV}$).
The rich set of final states in $e^{+}e^{-}$ and 
$\gamma\gamma$ collisions would play an essential role in measuring the mass,
spin, parity, two-photon width and trilinear self-coupling of the Higgs boson,
as well as its couplings to fermions and gauge bosons; these quantities are 
difficult to determine with only one initial state. All the measurements made
at LEP and SLC could be repeated using highly polarized electron beams and at
much higher luminosities. For some processes within and beyond the Standard
Model, the required CM energy is considerably lower at the proposed facility
than at an $e^{+}e^{-}$ or proton collider.} 
\end{minipage}
\end{center}

\vspace*{0.3cm}
\renewcommand{\thesection}{\arabic{section}}
\section{~Standard Model and Higgs mechanism}
\vspace*{0.3cm}

\setcounter{equation}{0}

~~~~Enormous progress has been made in the field of high-energy
physics over the past four decades. The existence of a subnuclear
world of quarks and leptons, whose dynamics can be described by quantum field
theories possessing gauge symmetry ({\em gauge theories}), has been firmly
established. The {\small\bf Standard Model} (SM) of particle physics gives a
coherent quantum-mechanical description of electromagnetic, weak and strong
interactions based on fundamental constituents --- quarks and leptons ---
interacting via force carriers --- photons, W and Z bosons, and gluons.

\begin{figure}[h]
\begin{center}
\epsfig{file=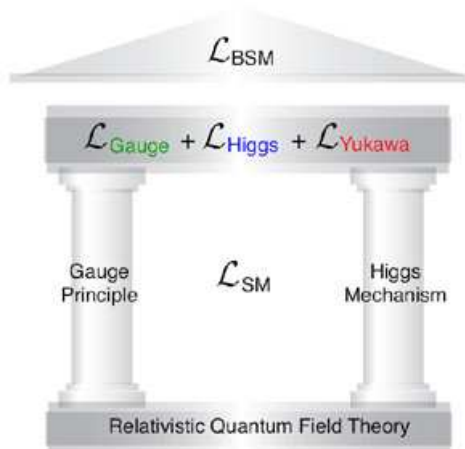,height=0.25\textheight}
\end{center}
\vskip -7mm
\caption{Two main pillars of the Standard Model (SM). The `BSM roof' represents
physics beyond the Standard Model. Reproduced courtesy of K. Fujii.} 
\label{fig:pillars}
\end{figure}

The Standard Model is supported by two theoretical `pillars': the {\small\bf
gauge principle} and the {\small\bf Higgs mechanism} for particle mass 
generation (see Fig.\,\ref{fig:pillars}). Whereas the former has been 
established through precision electroweak measurements, the latter
is essentially untested. 

In the SM, where electroweak symmetry is broken by the Higgs
mechanism, the mass of a particle depends on its interaction
with the Higgs field, a medium that permeates the universe. The photon and
gluon do not have such couplings, and so they remain massless. The Standard 
Model predicts the existence of a neutral spin-0 particle associated with the
Higgs field, but it does not predict its mass. Although the existence of a
Higgs field provides a simple mechanism for electroweak symmetry breaking,
{\em our inability to predict the mass of the Higgs boson reflects the fact
that we really do not understand at a fundamental level why this phenomenon
occurs}. Another undesirable feature of the Standard Model is the {\em ad hoc}
way in which fermion masses are introduced.

\begin{figure}[t]
\vspace{-3.3mm}
\begin{center}
\epsfig{file=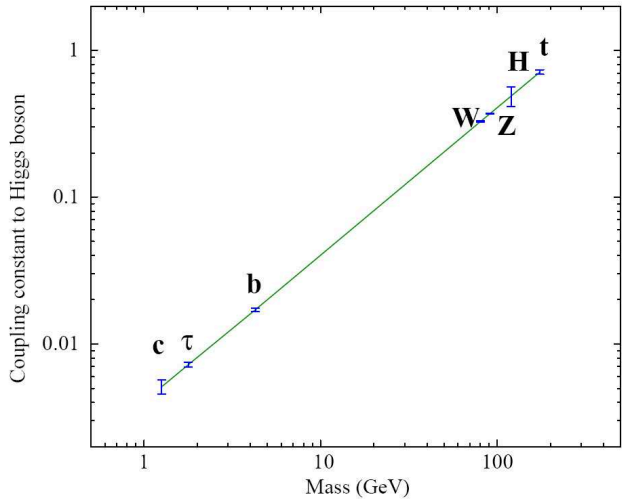,height=0.27\textheight}
\end{center}
\vskip -8mm
\caption{Precision with which the couplings of the Higgs particle with
$\mbox{\large{$m$}}_{\mbox{\tiny{H}}}^{~} = 120$ GeV can be determined at an
$e^{+}e^{-}$ collider with $\int\!L = 500~{\rm fb}^{-1}$. The coupling 
$\kappa_{i}$ of the particle $i$ with mass $m_{i}$ is defined so that the 
relation $m_{i} = v\kappa_{i}$ with $v \simeq 246$ GeV holds in the SM
\cite{GLC}.}
\label{fig:Higgs_couplings}
\end{figure}

All of the couplings of the Higgs particle to gauge bosons and fermions are
completely determined in the Standard Model in terms of electroweak coupling 
constants and fermion masses. Higgs production and decay processes can be 
computed in the SM unambiguously in terms of the Higgs mass alone. Since the
coupling of the Higgs boson to fermions and gauge bosons is proportional to the
particle masses (see Fig.\,\ref{fig:Higgs_couplings}), the Higgs boson will be
produced in association with heavy particles and will decay into the heaviest
particles that are kinematically accessible.
 
The rich set of final states in $e^{+}e^{-}$ and $\gamma\gamma$ collisions 
would play an essential role in measuring the mass ($\mbox{\large{$m$}}_{\mbox
{\tiny{H}}}^{~}$), spin, parity, two-photon width and trilinear self-coupling
of the Higgs boson, as well as its couplings to fermions and gauge bosons;
these quantities are difficult to determine with only one initial state.

The Higgs-boson mass affects the values of electroweak observables 
through radiative corrections. The precision electroweak data obtained over the
past two decades consists of over a thousand individual measurements. Many of 
those measurements may be combined to provide a global test of consistency with
the SM. The best constraint on $\mbox{\large{$m$}}_{\mbox{\tiny{H}}}^{~}$ is
obtained by making a global fit to the electroweak data. Such a fit strongly 
suggests that the most likely mass for the SM Higgs boson is just above the 
limit of 114.4 GeV set by direct searches at the LEP $e^{+}e^{-}$ collider
\cite{LEP}.
 
\begin{figure}[t]
\begin{center}
\epsfig{file=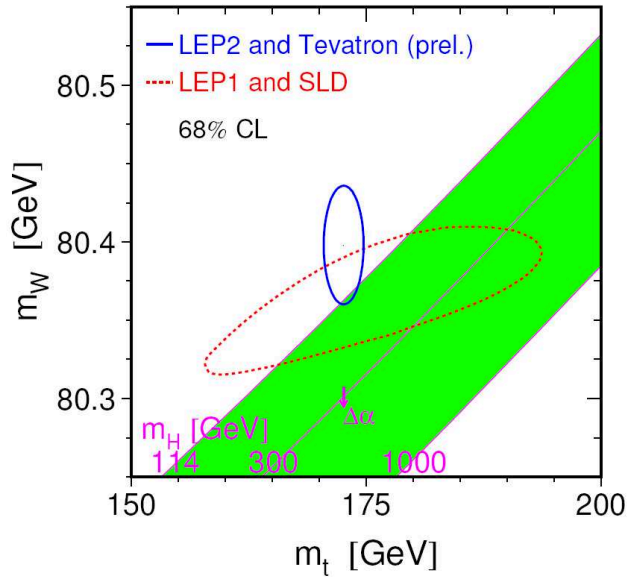,height=0.31\textheight}
\end{center}
\vskip -8mm
\caption{Direct and indirect constraints on the W and top-quark masses.
Reprinted courtesy of LEP Electroweak Working Group (March 2008).}
\label{fig:Mt-Mw}
\end{figure}

The dashed ellipse in Fig.\,\ref{fig:Mt-Mw} indicates the direct measurement of
the W mass, $\mbox{\large{$m$}}_{\mbox{\tiny{W}}}^{~}$, and the top-quark mass,
$\mbox{\large{$m$}}_{t}^{~}$. The elongated ellipse represents the predicted
relationship between the two masses. Also shown is the correlation between
$\mbox{\large{$m$}}_{\mbox{\tiny{W}}}^{~}$ and $\mbox{\large{$m$}}_{t}^{~}$ as
expected in the Standard Model for different values of the Higgs-boson mass
$\mbox{\large{$m$}}_{\mbox{\tiny{H}}}^{~}$ (the diagonal band). Notice that the
two ellipses overlap near the lines of constant $\mbox{\large{$m$}}_{\mbox
{\tiny{H}}}^{~}$. This indicates that the Standard Model is a fairly good
approximation to reality. Both ellipses are consistent with a low value of the
Higgs-boson mass.

High-precision electroweak measurements, therefore, provide a natural 
complement to direct studies of the Higgs sector. All the measurements made at
LEP and SLC could be repeated at the proposed facility using 90\% polarized
electron beams and at much higher luminosities. Assuming a geometric 
luminosity $L_{e^{+}e^{-}}^{~} \approx 5\times 10^{33}$ cm$^{-2}$\,s$^{-1}$ at
the Z resonance, about $2\times 10^{9}$ Z bosons can be produced in an 
operational year of $10^{7}$ s. This is about 200 times the entire LEP 
statistics. Moreover, about $10^{6}$ W bosons can be produced near the W-pair
threshold at the optimal energy point for measuring the W-boson mass. An 
increase in the number of Z events by two orders of magnitude as compared to 
LEP data, and a substantially improved accuracy in the measurement of W-boson
properties, would provide new opportunities for high-precision electroweak 
studies \cite{erler}.

\vspace*{0.3cm}
\renewcommand{\thesection}{\arabic{section}}
\section{~Higgs self-couplings}
\vspace*{0.3cm}

~~~~In order to provide a mechanism for the generation of particle masses in
the {\small\bf Standard Model} without violating its gauge invariance, a
complex scalar SU(2) doublet $\Phi$ with four real fields and {\em hypercharge}
${\rm Y} = 1$ is introduced. The dynamics of the field $\Phi$ is described by
the Lagrangian
\begin{equation}
{\cal L}_{\Phi} \,=\, (D_{\mu}\Phi )^{\dag}(D^{\mu}\Phi ) \,-\, \mu^{2\,}
\Phi^{\dag}\Phi \,-\, \lambda\mbox{\large{$($}}\Phi^{\dag}\Phi\mbox{\large
{$)$}}^{2}
\end{equation}
where $(D_{\mu}\Phi )^{\dag}(D^{\mu}\Phi )$ is the kinetic-energy term and 
$\mu^{2\,}\Phi^{\dag}\Phi + \lambda\mbox{\large{$($}}\Phi^{\dag}\Phi\mbox
{\large{$)$}}^{2}$ is the Higgs self-interaction potential. In the so-called
{\em unitary gauge},
\begin{equation}
\Phi \,=\, \frac{\raisebox{-.3ex}{\mbox{\small{1}}}}{\sqrt{\mbox{\small
{2}}}}\left (\!\!
\begin{array}{c}
0 \\*[0.7mm]
v + {\rm H}
\end{array} \!\right )
\end{equation}
where $v \equiv \sqrt{-\mu^{2}/\lambda} = 246~{\rm GeV}$ is the {\small\bf
vacuum expectation value} of the scalar field $\Phi$. The Higgs 
self-interaction potential gives rise to terms involving only the physical
{\small\bf Higgs field} H:
\begin{equation}
V_{\mbox{\tiny{H}}} \,=\, \frac{\raisebox{-.3ex}{\mbox{\small{1}}}}
{\raisebox{.2ex}{\mbox{\small{2}}}}\mbox{\Large{$($}}\mbox{\small{2}}\lambda
v^{2}\mbox{\Large{$)$}}{\rm H}^{2} \,+\, \lambda v\,{\rm H}^{3} \,+\, \frac
{\raisebox{-.3ex}{$\lambda$}}{\raisebox{.2ex}{\mbox{\small{4}}}}\,{\rm H}^{4}
\end{equation}

We see from Eq. (3) that the {\small\bf Higgs mass} $\mbox{\large{$m$}}_{\mbox
{\tiny{H}}}^{~} \,=\, \sqrt{\mbox{\small{2}}\lambda}\,v$ is related to the
quadrilinear self-coupling strength $\lambda$. It is also evident that the
{\small\bf trilinear self-coupling} of the Higgs field is 
\begin{equation}
\lambda_{\mbox{\tiny{HHH}}} \,\equiv\, \lambda v \,=\, \frac{\mbox{\large{$m$}}
_{\mbox{\tiny{H}}}^{\,2}}{\raisebox{.3ex}{\mbox{\small{2}}$v$}}
\label{eq:hhh}
\end{equation}
and the self-coupling among four Higgs fields
\begin{equation}
\lambda_{\mbox{\tiny{HHHH}}} \,\equiv\, \frac{\raisebox{-.3ex}{$\lambda$}}
{\raisebox{.2ex}{\mbox{\small{4}}}} \,=\, \frac{\mbox{\large{$m$}}_{\mbox{\tiny
{H}}}^{\,2}}{\mbox{\small{8}}v^{2}}
\end{equation}
Note that the Higgs self-couplings are uniquely determined by the mass of the 
Higgs boson, which represents a free parameter of the model.

Any theoretical model based on the gauge principle must evoke spontaneous
symmetry breaking. For example, the minimal {\em supersymmetric} extension of
the Standard Model ({\small\bf MSSM}) introduces two SU(2) doublets of complex
Higgs fields, whose neutral components have vacuum expectation values 
$v_{\mbox{\tiny{1}}}$ and $v_{\mbox{\tiny{2}}}$. In this model,
spontaneous electroweak symmetry breaking results in five physical
Higgs-boson states: two neutral scalar fields $h^0$ and $H^0$, a
pseudoscalar $A^0$ and two charged bosons $H^\pm$. This extended Higgs
system can be described at tree level by two parameters: the ratio
$\tan\beta \equiv v_{\mbox{\tiny{2}}}/v_{\mbox{\tiny{1}}}$, and a mass
parameter, which is generally identified with the mass of the
pseudoscalar boson $A^0$, $\mbox{\large{$m$}}_{\mbox{\tiny{A}}}^{~}$. While 
there is a bound of about 140~GeV on the mass of the lightest CP-even neutral
Higgs boson $h^0$ \cite{h0-mass1,h0-mass2}, the masses of the $H^0$, $A^0$ and
$H^\pm$ bosons may be much larger. The existence of the Higgs boson $h^0$
is the only verifiable low-energy prediction of the MSSM model.

The trilinear self-coupling of the lightest MSSM Higgs boson at tree level is
given by
\begin{equation}
\lambda_{hhh}  \,=\, \frac{\mbox{\large{$m$}}_{\mbox{\tiny{Z}}}^{\,2}}
{\raisebox{.3ex}{\mbox{\small{2}}$v$}}\cos 2\alpha\sin(\beta+\alpha)
\end{equation}
where
\begin{equation}
\tan 2\alpha \,=\, \tan 2\beta \,\frac{\mbox{\large{$m$}}_{\mbox{\tiny{$A$}}}^
{\,2} + \mbox{\large{$m$}}_{\mbox{\tiny{Z}}}^{\,2}}{\mbox{\large{$m$}}_{\mbox
{\tiny{$A$}}}^{\,2} - \mbox{\large{$m$}}_{\mbox{\tiny{Z}}}^{\,2}}
\end{equation}
We see that for arbitrary values of the MSSM input parameters $\tan\beta$ and
$\mbox{\large{$m$}}_{\mbox{\tiny{$A$}}}^{~}$ the value of the $h^0$
self-coupling differs from that of the SM Higgs boson. However, in the
so-called `decoupling limit' $\mbox{\large{$m$}}_{\mbox{\tiny{$A$}}}^{\,2} \sim
\mbox{\large{$m$}}_{\mbox{\tiny{$H$}}^{0}}^{\,2} \sim \mbox{\large{$m$}}_{\mbox
{\tiny{$H$}}^{\pm}}^{\,2} \gg v^2/2$, the trilinear and quadrilinear 
self-couplings of the lightest CP-even neutral Higgs boson $h^0$ approach the
SM value. 

In contrast to any anomalous couplings of the gauge bosons, an anomalous 
self-coupling of the Higgs particle would contribute to electroweak 
observables only at two-loop and higher orders, and is therefore practically 
unconstrained by precision electroweak measurements \cite{vanderBij:1985ww}.

\vspace*{0.3cm}
\renewcommand{\thesection}{\arabic{section}}
\section{~Single Higgs production in $\gamma\gamma$ collisions}
\vspace*{0.3cm}

~~~~Since photons couple directly to all fundamental fields carrying the
electromagnetic current (leptons, quarks, W bosons, supersymmetric particles),
$\gamma\gamma$ collisions provide a comprehensive means of exploring virtually
every aspect of the SM and its extensions (see \cite{boos}, \cite{belusev} and
references therein). The production mechanisms in $e^{+}e^{-}$ collisions are
often more complex and model-dependent. Moreover, the cross-sections for
production of charged-particle pairs in $\gamma\gamma$ interactions are 
approximately an order of magnitude larger than in $e^{+}e^{-}$ annihilations.

In $\gamma\gamma$ collisions, the {\small\bf Higgs boson} is produced as a 
single resonance in a state of definite CP, which is perhaps the most important
advantage over $e^{+}e^{-}$ annihilations, where this $s$-channel process is
highly suppressed. For the Higgs mass in the range $\mbox{\large{$m$}}_{\mbox
{\tiny{H}}}^{~} =$ 115$-$200 GeV, the effective cross-section for $\gamma\gamma
\rightarrow$ H is about a factor of five larger than that for Higgs production
in $e^{+}e^{-}$ annihilations. In this mass range, the process $e^{+}e^{-}
\rightarrow {\rm HZ}$ requires considerably higher center-of-mass energies than
$\gamma\gamma \rightarrow$ H. In $e^{+}e^{-}$ annihilations, the heavy neutral
MSSM Higgs bosons can be created only by associated production ($e^{+}e^{-}
\rightarrow H^{0}A^{0}$), whereas in $\gamma\gamma$ collisions they are
produced as single resonances ($\gamma\gamma \rightarrow H^{0},\,A^{0}$) with
masses up to 80\% of the initial $e^{-}e^{-}$  collider energy \cite{zerwas}.

Furthermore, calculations show that the {\em statistical} sensitivity of the 
cross-section $\mbox{\large{$\sigma$}}_{\gamma\gamma\,\rightarrow\,\mbox{\tiny
{HH}}}$ to the {\small\bf Higgs self-coupling} is maximal near the 
$2\mbox{\large{$m$}}_{\mbox{\tiny{H}}}^{~}$
threshold for $\mbox{\large{$m$}}_{\mbox{\tiny{H}}}^{~}$
between 115 and 160 GeV, and is comparable with the statistical sensitivities 
of $\mbox{\large{$\sigma$}}_{e^{+}e^{-}\,\rightarrow\,\mbox{\tiny{HHZ}}}$ and 
$\mbox{\large{$\sigma$}}_{e^{+}e^{-}\,\rightarrow\,\mbox{\tiny{HH}}\nu\bar
{\nu}}$ to this coupling for ${\rm E}_{ee} \leq 700$ GeV \cite{belusev1}. The
overall {\em acceptance} is expected to be considerably larger in $\gamma\gamma
\rightarrow$ HH than in the process $e^{+}e^{-} \rightarrow {\rm HH}\nu\bar
{\nu}$. Note also that hadron colliders are not well suited for measuring the
self-coupling of the Higgs boson if $\mbox{\large{$m$}}_{\mbox{\tiny{H}}} \leq
140$ GeV \cite{baur}.

The reaction $\gamma\gamma \rightarrow$ H, which is related to ${\rm H}
\rightarrow \gamma\gamma$, proceeds through a `loop diagram' and receives
contributions from {\em all} charged particles that couple to the photon and
the Higgs boson. Thus, the {\small\bf two-photon width} $\Gamma ({\rm H}
\rightarrow \gamma\gamma )$ is sensitive to the Higgs-top Yukawa coupling, as
well as mass scales far beyond the energy of the $\gamma\gamma$ collision.
Assuming that the branching ratio ${\rm BR}({\rm H} \rightarrow b\bar{b})$ can
be measured to an accuracy of about 2\% in the process $e^{+}e^{-} \rightarrow
{\rm HZ}$, the $\gamma\gamma$ partial width can be determined with a similar 
precision for $\mbox{\large{$m$}}_{\mbox{\tiny{H}}}^{~} \simeq 120$ GeV by
measuring the cross-section $\mbox{\large{$\sigma$}}(\gamma\gamma \rightarrow
{\rm H} \rightarrow b\bar{b}) \propto \Gamma ({\rm H} \rightarrow \gamma\gamma)
{\rm BR}({\rm H} \rightarrow b\bar{b})$. The Higgs-top coupling can also be 
measured in the process $e^{+}e^{-} \rightarrow t\bar{t}$ at the 
pair-production threshold \cite{fujii}.

\begin{figure}[t]
\begin{center}
\epsfig{file=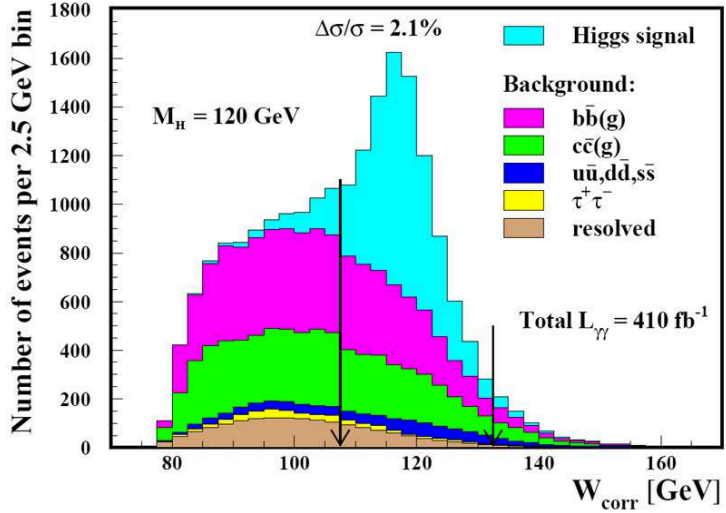,height=0.28\textheight}
\end{center}
\vskip -6mm
\caption{The reconstructed invariant-mass distribution of the $\gamma\gamma
\rightarrow {\rm H} \rightarrow b\bar{b}$ signal and the $b\bar{b}(g)$ and
$c\bar{c}(g)$ backgrounds. The gluon (`resolved') structure of the photon can 
be measured {\em in situ}. Credit: P. Niezurawski, A. Zarnecki and
M. Krawczyk.}
\label{fig:invariant_mass}
\end{figure}

Both the energy spectrum and polarization of the backscattered photons depend
strongly on the polarizations of the incident electrons and laser photons. The
key advantage of using $e^{-}e^{-}$ beams is that they can be polarized to a
high degree, enabling one to tailor the photon energy distribution to one's
needs. In a collision of two photons, the possible helicities are 0 or 2. For
example, the Higgs boson is produced in the $J_{z} = 0$ state, whereas 
the background processes $\gamma\gamma \rightarrow b\bar{b},\,c\bar{c}$ are
suppressed for this helicity configuration 
(see Fig.\,\ref{fig:invariant_mass}).
The circular polarization of the photon beams is therefore an important asset,
for it can be used both to enhance the signal and suppress the background.

The {\small\bf CP properties} of any neutral Higgs boson that may be produced 
at a photon collider can be {\em directly} determined by controlling the 
polarizations of Compton-scattered photons \cite{grzadkowski}. A CP-even Higgs
boson couples to the combination ${\bf e}_{\mbox{\tiny{1}}}\mbox{\boldmath
{$\cdot$}}\,{\bf e}_{\mbox{\tiny{2}}}$, whereas a CP-odd Higgs boson couples to
$({\bf e}_{\mbox{\tiny{1}}}\!\times\!{\bf e}_{\mbox{\tiny{2}}})\,\mbox{\boldmath
{$\cdot$}}\,\mbox{\boldmath{$k$}}_{\gamma}$:
\[ {\cal M}(\gamma\gamma\rightarrow{\rm H}[0^{++}]) \,\propto\, {\bf e}_{\mbox
{\tiny{1}}}\mbox{\boldmath{$\cdot$}}\,{\bf e}_{\mbox{\tiny{2}}} \,\propto\,
1 + \cos 2\phi \]
\[ {\cal M}(\gamma\gamma\rightarrow A[0^{-+}]) \,\propto\, ({\bf e}_{\mbox
{\tiny{1}}}\!\times\!{\bf e}_{\mbox{\tiny{2}}})\,\mbox{\boldmath{$\cdot$}}\,
\mbox{\boldmath{$k$}}_{\gamma} \,\propto\, 1 - \cos 2\phi \] 
where ${\bf e}_{i}^{~}$ are polarization vectors of colliding photons, $\phi$ is
the angle between them, and $\mbox{\boldmath{$k$}}_{\gamma}$ is the momentum
vector of one of the Compton-scattered photons; $0^{++}$ and $0^{-+}$ are the
quantum numbers ${\rm J}^{\rm PC}$. The scalar (pseudoscalar) Higgs boson 
couples to {\em linearly polarized} photons with a maximum strength if the 
polarization vectors are parallel (perpendicular): $\mbox{\large{$\sigma$}}
 \propto 1 \pm l_{\mbox{\tiny{1}}}l_{\mbox{\tiny{2}}}\cos 2\phi$, where $l_{i}$
are the degrees of linear polarization; the signs $\pm$ correspond to the
${\rm CP} = \pm 1$ particles. 

The general amplitude for a CP-{\em mixed state} to couple to the two photons is
\begin{equation}
{\cal M} \,=\, {\cal E}({\bf e}_{\mbox{\tiny{1}}}\mbox{\boldmath{$\cdot$}}\,
{\bf e}_{\mbox{\tiny{2}}}) \,+\, {\cal O}({\bf e}_{\mbox{\tiny{1}}}\!\times\!
{\bf e}_{\mbox{\tiny{2}}})_{z}^{~}
\end{equation}
where ${\cal E}$ is the CP-even and ${\cal O}$ the CP-odd contribution to the 
amplitude. If we denote the {\em helicities} of the two photons by $\lambda_{1}$
and $\lambda_{2}$, with $\lambda_{1},\lambda_{2} = \pm 1$, then the above
vector products can be expressed as 
\[ {\bf e}_{\mbox{\tiny{1}}}\mbox{\boldmath{$\cdot$}}\,{\bf e}_{\mbox{\tiny{2}}}
 \,=\, -(1 + \lambda_{1}\lambda_{2})/2~~~~~~~~~~~~~~~
({\bf e}_{\mbox{\tiny{1}}}\!\times\!{\bf e}_{\mbox{\tiny{2}}})_{z}^{~} \,=\,
 i\lambda_{1}(1 + \lambda_{1}\lambda_{2})/2 \]
Now, $|{\cal M}_{++}|^{2} - |{\cal M}_{--}|^{2} = -4{\rm Im}({\cal E}{\cal O}
^{*}),~~2{\rm Re}({\cal M}_{--}^{*}{\cal M}_{++}) = 2(|{\cal E}|^{2} -|{\cal O}|
^{2})$ and $2{\rm Im}({\cal M}_{--}^{*}{\cal M}_{++}) = -4{\rm Re}({\cal E}
{\cal O}^{*})$. When these expressions are divided by $|{\cal M}_{++}|^{2} + 
|{\cal M}_{--}|^{2} = 2(|{\cal E}|^{2} + |{\cal O}|^{2})$, we obtain three
polarization asymmetries that give an unambiguous measure of CP-mixing
\cite{grzadkowski}. Note that these require both {\em linearly} and 
{\em circularly} polarized photons.

In $e^{+}e^{-}$ annihilations, it is straightforward to discriminate between
CP-even and CP-odd neutral Higgs bosons, but would be difficult to detect small
CP-violating effects (which contribute only at the one-loop level) for a
dominantly CP-even component (which contributes at the tree level in 
$e^{+}e^{-}$ collisions) \cite{hagiwara}.

\vspace*{0.3cm}
\section{~Single Higgs production in $e^{+}e^{-}$ annihilations}
\vspace*{0.3cm}

\begin{figure}[h] 
\begin{center}
\epsfig{file=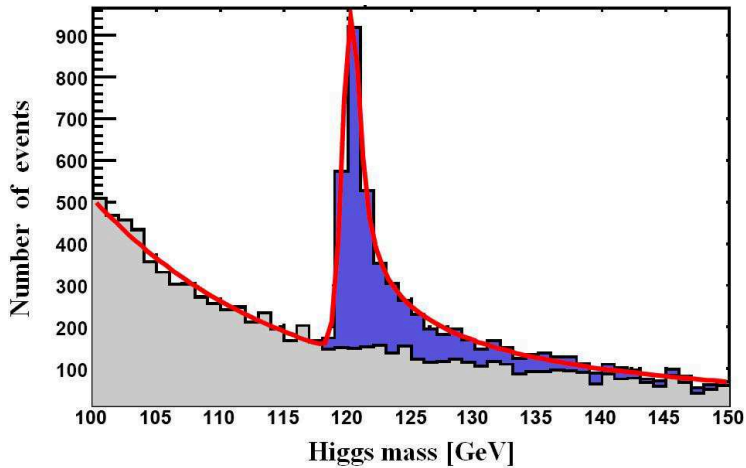,height=0.25\textheight}
\end{center}
\vskip -6mm
\caption{Distribution of the invariant mass of the system recoiling against a
pair of leptons in the process $e^{+}e^{-} \rightarrow {\rm HZ} \rightarrow
X\ell^{+}\ell^{-}$ for $\mbox{\large{$m$}}_{\mbox{\tiny{H}}}^{~} = 120$ GeV and
$\int\!{\cal L} = 500\,fb^{-1}$ at $\sqrt{s} = 250$ GeV. The red line is a fit 
to a Monte Carlo simulation of the Higgs signal and the ZZ background; the gray
area represents the background only \cite{Abe}. For $\mbox{\large{$m$}}_{\mbox
{\tiny{H}}}^{~} \simeq 120$ GeV, the optimum center-of-mass energy is $\sqrt{s}
\simeq 230$ GeV.}
\label{fig:recoil_mass}
\end{figure}

~~~~A particularly noteworthy feature of an $e^{+}e^{-}$ collider is that
the Higgs boson can be detected in the {\small\bf Higgs-strahlung process}
\begin{equation}
e^{+}e^{-} \,\rightarrow\, {\rm HZ}
\end{equation}
even if it decays into invisible particles (e.g., the lightest {\em neutralino}
of a supersymmetric model). In this case the signal manifests itself as a peak
in the distribution of invariant mass of the system recoiling against the 
lepton pair stemming from Z-boson decay (see Fig.\,\ref{fig:recoil_mass}).

By exploiting the ${\rm HZ} \rightarrow X\ell^{+}\ell^{-}$ channel, the
Higgs-strahlung {\small\bf cross-sections} can be measured with statistical
errors of 2.6 to 3.1 percent for Higgs-boson masses from 120 to 160 GeV (see
\cite{heinemeyer} and references therein).

From the fits to the reconstructed mass spectra in the channels ${\rm HZ} 
\rightarrow q\bar{q}\ell^{+}\ell^{-},~b\bar{b}q\bar{q},~{\rm WW}\ell^{+}
\ell^{-}$ and ${\rm WW}q\bar{q}$, the {\small\bf Higgs-boson mass} can be 
determined with an uncertainty of 40 to 70 MeV for $\mbox{\large{$m$}}_{\mbox
{\tiny{H}}}^{~}$ in the range 120 to 180 GeV \cite{heinemeyer}.

To determine the {\small\bf spin} and {\small\bf parity} of the SM Higgs boson
in the Higgs-strahlung process, one can use the information on (1) the energy
dependence of the Higgs-boson production cross-section just above the kinematic
threshold, and (2) the angular distribution of the Z/H bosons. The best way to
study the {\small\bf CP properties} of the Higgs boson is by analyzing the spin
correlation effects in the decay channel ${\rm H} \rightarrow \tau^{+}\tau^{-}$
(see \cite{heinemeyer} and references therein). 
 
The Higgs-strahlung cross-section, which dominates at low CM energies,
decreases with energy in proportion to $1/s$ (see Eq. (21)). In contrast, the 
cross-section for the {\small\bf W-fusion process}
\begin{equation}
e^{+}e^{-} \,\rightarrow\, {\rm H}\nu_{e}^{~}\bar{\nu}_{e}^{~}
\end{equation}
increases with energy in proportion to log($s/\mbox{\large{$m$}}_{\mbox{\tiny
{H}}}^{\,2}$), and hence becomes more important at energies $\sqrt{s} \gsim 
500$ GeV for the Higgs-mass range $115~{\rm GeV} \lsim \mbox{\large{$m$}}_
{\mbox{\tiny{H}}}^{~} \lsim 200~{\rm GeV}$.

\vspace*{0.3cm}
\section{~Higgs-pair production in $\gamma\gamma$ and $e^+e^-$ collisions}
\vspace*{0.3cm}

~~~~It is well known that hadron colliders are not ideally suited for measuring
the self-coupling of the Higgs boson if $\mbox{\large{$m$}}_{\mbox{\tiny{H}}}
^{~}\,\leq\,140$~GeV \cite{baur}. The potential of a future $\gamma\gamma /
e^+e^-$ collider for determining the HHH coupling has therefore been closely 
examined (see \cite{belusev1} and
\cite{Djouadi:1999gv,Miller:1999ct,Castanier:2001sf,Belanger:2003ya}).

\begin{figure}[t]
\vskip -3mm
\setlength{\unitlength}{1cm}
\begin{picture}(10,10)
\put(0.25,0){\epsfig{file=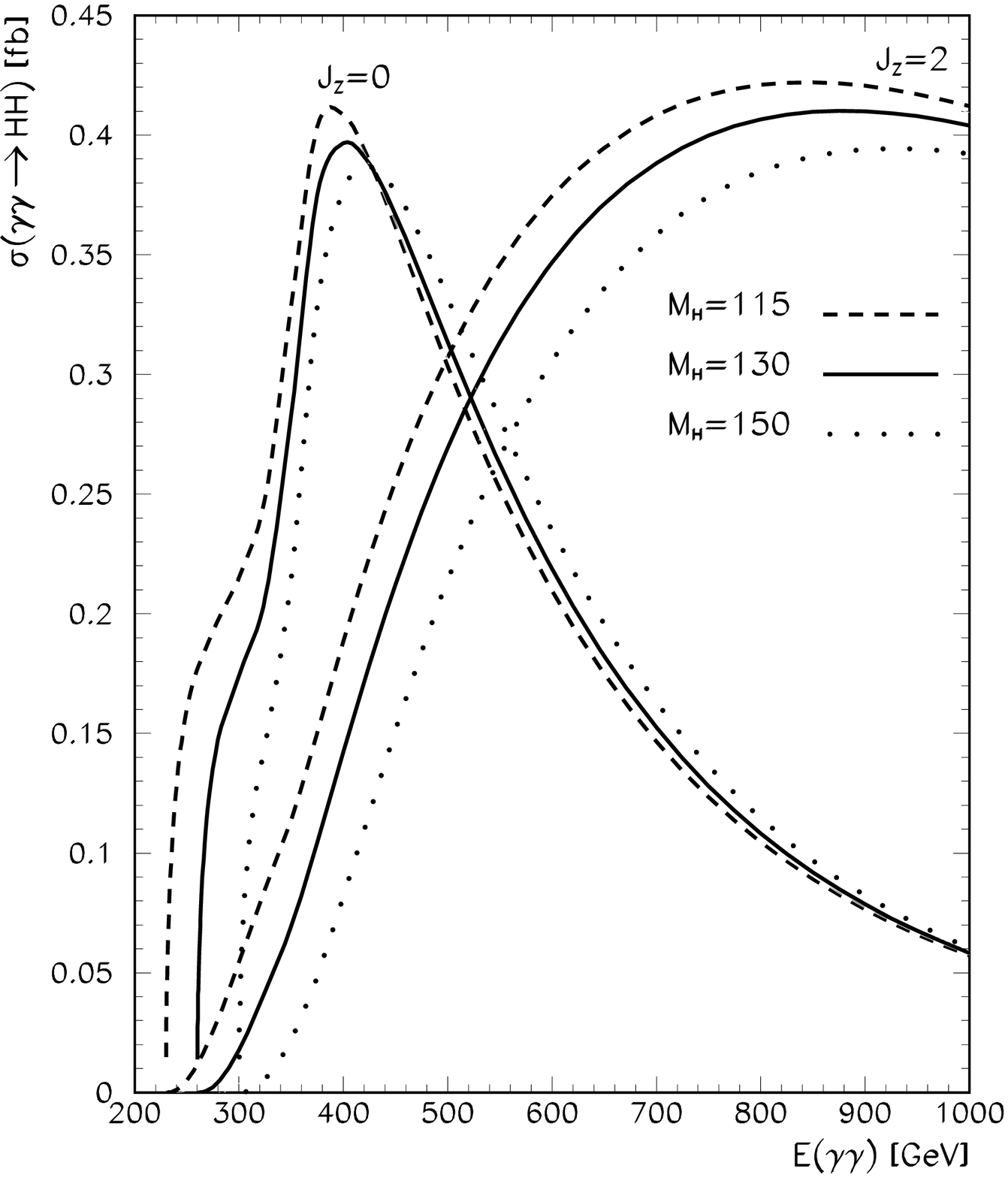,width=0.5\textwidth}}
\put(8.5,0){\epsfig{file=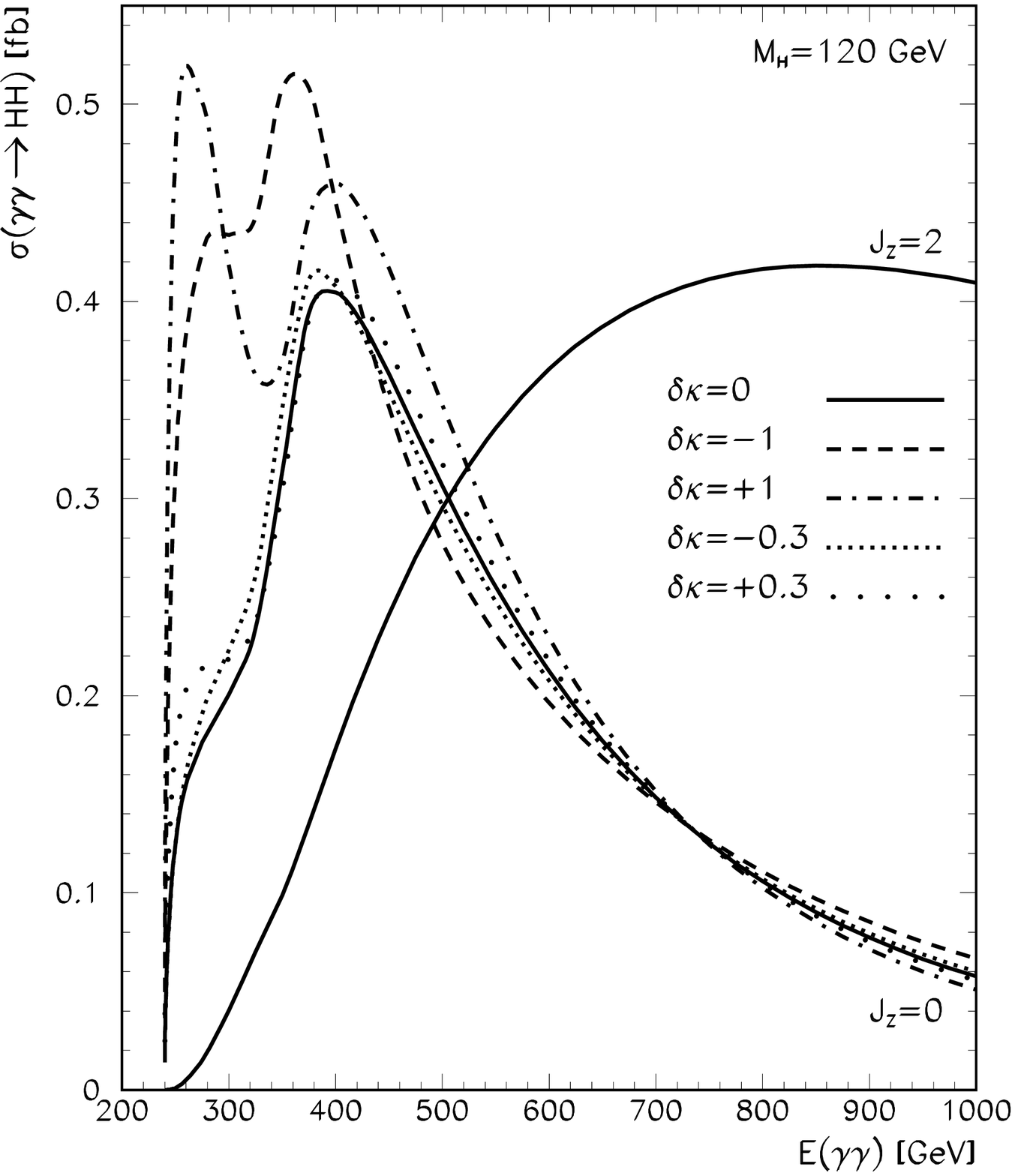,width=0.5\textwidth}}
\end{picture}
\vskip -8mm
\caption{(a) The total $\gamma\gamma\to{\rm HH}$ cross-section as a
  function of the $\gamma\gamma$ center-of-mass energy. Contributions
  for equal ($J_z=0$) and opposite ($J_z=2$) photon helicities are
  shown separately.
  \newline
  (b) The cross-sections for HH production in $\gamma\gamma$ collisions for
$\mbox{\large{$m$}}_{\mbox{\tiny{H}}}^{~} = 120$ GeV and anomalous trilinear 
Higgs self-couplings $\delta\kappa=0,\pm 1, \pm 0.3$. Credit: R. Belusevic \& 
G. Jikia \cite{belusev1}.}
\label{fig:gghh}
\end{figure}

The production of a pair of SM Higgs bosons in photon-photon collisions,
\begin{equation}
\gamma\gamma \,\to\, {\rm HH}
\end{equation}
which is related to the Higgs-boson decay into two
photons, is due to W-boson and top-quark box and triangle loop diagrams. The
total cross-section for $\gamma\gamma\to{\rm HH}$ in polarized photon-photon
collisions, calculated at the leading one-loop order \cite{Jikia:1992mt} as a
function of the $\gamma\gamma$ center-of-mass energy and for 
$\mbox{\large{$m$}}_{\mbox{\tiny{H}}}^{~}$
between 115 and 150 GeV, is shown in Fig.\,\ref{fig:gghh}a. The
cross-section calculated for equal photon helicities, $\mbox{\large{$\sigma$}}_
{\gamma\gamma\,\rightarrow\,\mbox{\tiny{HH}}}(\mbox{\small{$J_{z}=0$}})$, rises
sharply above the $2\mbox{\large{$m$}}_{\mbox{\tiny{H}}}^{~}$ 
threshold for different values of $\mbox{\large{$m$}}_{\mbox{\tiny{H}}}^{~}$,
and has a peak value of about $0.4$~fb at a 
$\gamma\gamma$ center-of-mass energy of 400~GeV. In contrast, the cross-section
for opposite photon helicities, $\mbox{\large{$\sigma$}}_{\gamma\gamma\,
\rightarrow\,\mbox{\tiny{HH}}}(\mbox{\small{$J_{z}=2$}})$, rises more slowly
with energy because a pair of Higgs bosons is produced in a state with orbital
angular momentum of at least $2\hbar$.

The cross-sections for equal photon helicities are of special interest, since
only the $J_z=0$ amplitudes contain contributions with trilinear Higgs 
self-coupling. By adding to the SM Higgs potential $V(\Phi^{\dag}\Phi )$ 
a gauge-invariant dimension-6 operator $\mbox{\large{$($}}\Phi^{\dag}\Phi\mbox
{\large{$)$}}^{3}$, one introduces a gauge-invariant anomalous trilinear
Higgs coupling $\delta\kappa$ \cite{Jikia:1992mt}. For the reaction $\gamma
\gamma \to {\rm HH}$, the only effect of such a coupling in the {\em unitary
gauge} would be to replace the trilinear HHH coupling of the SM in Eq.
\ref{eq:hhh} by an {\small\bf anomalous Higgs self-coupling} 
\begin{equation}
\lambda \,=\, (1 + \delta\kappa )\lambda_{\mbox{\tiny{HHH}}}^{~} 
\end{equation}
The dimensionless anomalous coupling $\delta\kappa$ is normalized so that 
$\delta\kappa=-1$ exactly cancels the SM HHH coupling. The cross-sections
$\mbox{\large{$\sigma$}}_{\gamma\gamma\,\rightarrow\,\mbox{\tiny{HH}}}$ for
five values of $\delta\kappa$ are shown in Fig.~\ref{fig:gghh}b.

In an experiment to measure the trilinear Higgs self-coupling, the contribution
from $\gamma\gamma \to {\rm HH}$ for opposite photon helicities represents
an irreducible background. However, this background is suppressed if one
chooses a $\gamma\gamma$ center-of-mass energy below about 320 GeV. 

To ascertain the potential of $\gamma\gamma$ colliders for measuring an 
anomalous trilinear Higgs self-coupling, one must take into account the fact 
that the photons are not monochromatic \cite{Ginzburg:1982yr}.
It is envisaged that an $e^{-}e^{-}$ linac and a terawatt laser system will be
used to produce Compton-scattered $\gamma$-ray beams for a photon collider.
Both the energy spectrum and polarization of the backscattered photons depend
strongly on the polarizations of the incident electrons and photons. A
longitudinal electron-beam polarization of 90\% and a 100\% circular
polarization of laser photons are assumed throughout.

\begin{figure}[t]
\vskip -3mm
\setlength{\unitlength}{1cm}
\begin{picture}(10,10)
\put(0.25,0){\epsfig{file=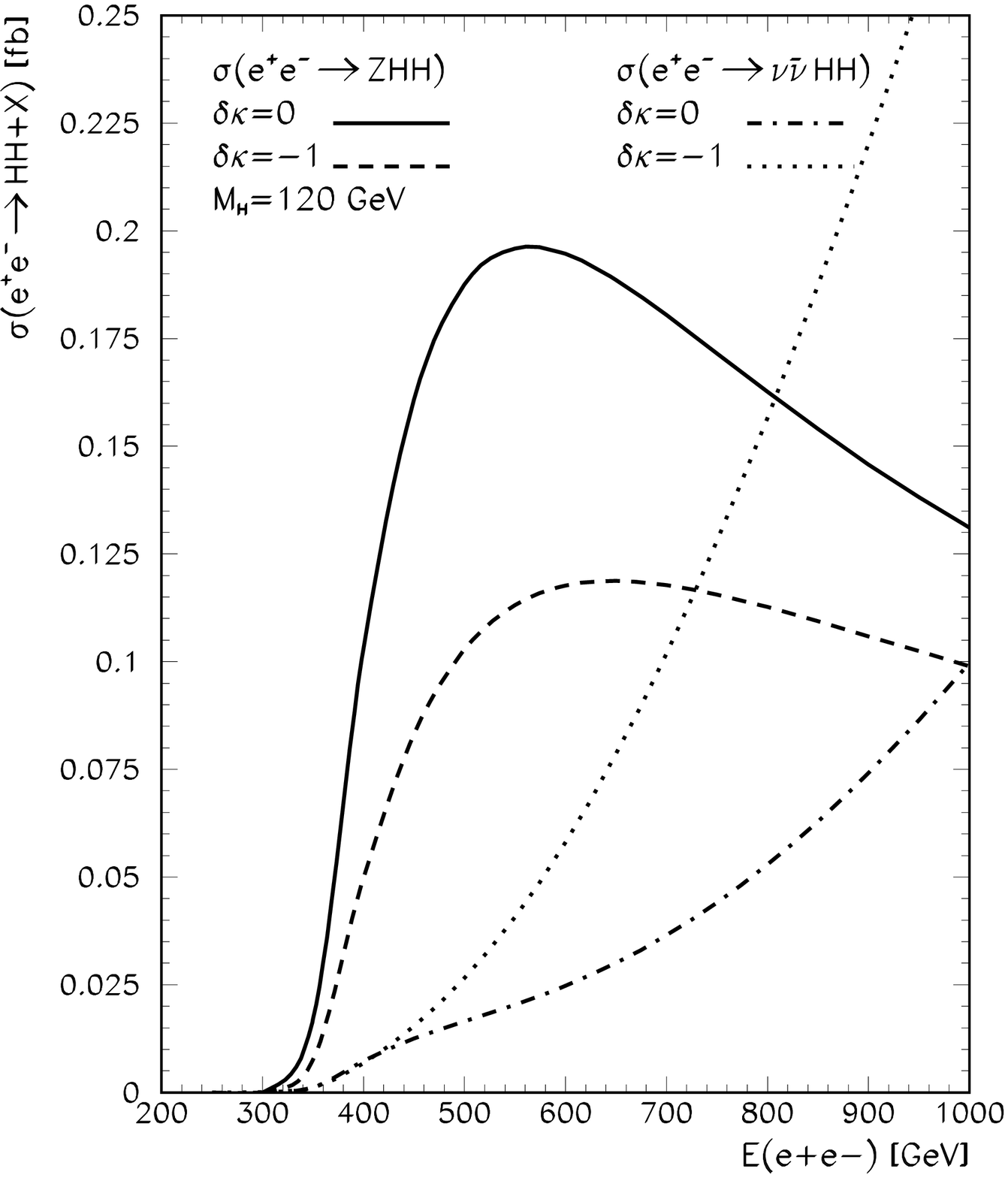,width=0.5\textwidth}}
\put(8.5,0){\epsfig{file=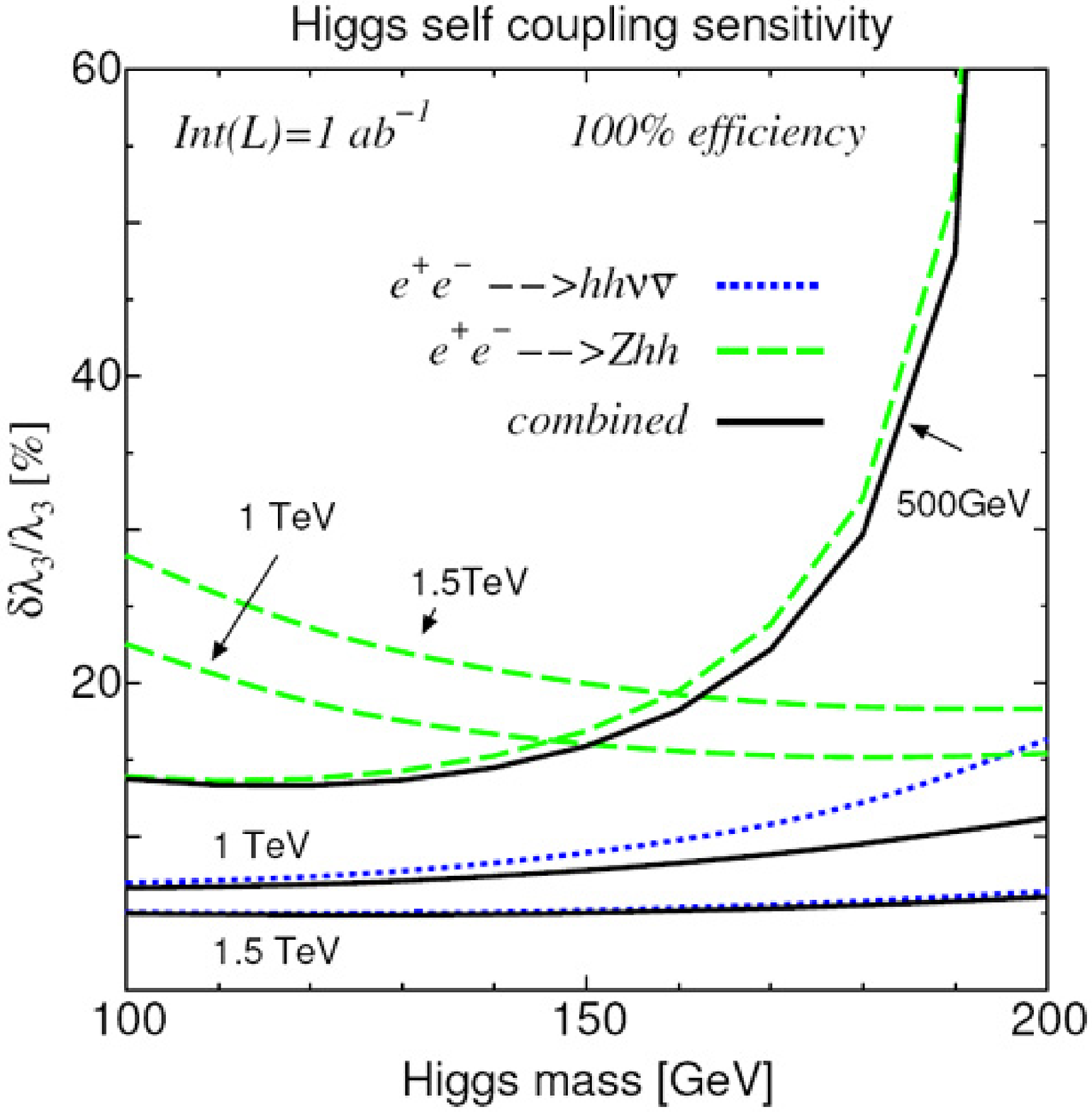,width=0.48\textwidth}}
\end{picture}
\vskip -1mm
\caption{(a) Total cross-sections for $e^+e^-\,\to\, {\rm HH}Z$ and 
$e^+e^-\,\to\,{\rm HH}\nu_e\bar\nu_e$ as functions of the $e^+e^-$
CM energy for $\mbox{\large{$m$}}_{\mbox{\tiny{H}}}^{~}=120$
GeV and the anomalous trilinear Higgs self-couplings $\delta\kappa=0$ or $-1$;
Credit: R. Belusevic \& G. Jikia \cite{belusev1}. (b) Statistical sensitivity 
of the trilinear self-coupling for the processes $e^+e^-\,\to\, {\rm HH}Z$ and
$e^+e^-\,\to\,{\rm HH}\nu_e\bar\nu_e$ \cite{GLC}.} 
\label{fig:eehh}
\end{figure}

The trilinear self-coupling of the Higgs boson can also be measured either in 
the so-called {\small\bf double Higgs-strahlung process}
\begin{equation}
e^+e^- \,\to\, {\rm HHZ}
\end{equation}
or in the {\small\bf W-fusion reaction}
\begin{equation}
e^+e^- \,\to\, {\rm HH}\nu_e^{~}\bar{\nu}_e^{~}
\end{equation}
The total cross-section for pair production of 120-GeV Higgs bosons in $e^+e^-$
collisions, calculated for {\em unpolarized} beams, is presented in
Fig.~\ref{fig:eehh} for anomalous trilinear Higgs self-couplings $\delta
\kappa=0$ or $-1$. If the electron beam is 100\% polarized, the double
Higgs-strahlung cross-section will stay approximately the same, while the
W-fusion cross-section will be twice as large. From Fig.~\ref{fig:eehh}
we infer that the SM double Higgs-strahlung cross-section exceeds 0.1~fb at 
400~GeV for $\mbox{\large{$m$}}_{\mbox{\tiny{H}}}^{~}\,=\,120$~GeV, and 
reaches a broad maximum of about 0.2~fb at a CM energy of 550~GeV. The SM 
cross-section for W-fusion stays below 0.1~fb for CM energies up to 1 TeV.  

For $\mbox{\large{$m$}}_{\mbox{\tiny{H}}}^{~}\,=\,120$~GeV, and assuming a
longitudinal electron-beam polarization of 90\%, the maximum sensitivity to an
anomalous trilinear Higgs self-coupling is achieved in the double 
Higgsstrahlung process at a CM energy of about 500~GeV \cite{belusev1}. This is
significantly higher than the optimal CM energy in $\gamma\gamma$ collisions.
In the W-fusion process, a similar sensitivity is attained at ${\rm E}(e^+e^-)
\approx 600$~GeV. 

Calculations show that the {\em statistical} sensitivity of
$\mbox{\large{$\sigma$}}_{\gamma\gamma\,\rightarrow\,\mbox{\tiny{HH}}}$ to the
Higgs self-coupling is maximal near the kinematic threshold for Higgs-pair
production for $\mbox{\large{$m$}}_{\mbox{\tiny{H}}}$ between 115 and 150~GeV,
and is comparable with the sensitivities of $\mbox{\large{$\sigma$}}_{e^{+}
e^{-}\,\rightarrow\,\mbox{\tiny{HHZ}}}$ and $\mbox{\large{$\sigma$}}_{e^{+}
e^{-}\,\rightarrow\,\mbox{\tiny{HH}}\nu\bar{\nu}}$ to this coupling for 
${\rm E}_{ee}\leq 700$ GeV, even if the integrated luminosity in $\gamma\gamma$
collisions is only one third of that in $e^+e^-$ annihilations \cite{belusev1}.
As mentioned earlier, the overall {\em acceptance} is expected to be 
considerably larger in the process $\gamma\gamma \rightarrow {\rm HH}$ than in
the reaction $e^{+}e^{-} \rightarrow {\rm HH}\nu\bar{\nu}$.

The Feynman diagrams for the process $\gamma\gamma \rightarrow {\rm HH}$ are
shown in Fig.~1 of \cite{Jikia:1992mt}. New physics beyond the Standard Model
introduces additional complexity into the subtle interplay between the
Higgs `pole amplitudes' and the top-quark and W-boson `box diagrams':
\[ |{\cal M}(J_{z} = 0)|^{2} \,=\, |A(s)(\lambda_{\mbox{\tiny{SM}}}^{~} +
\delta\lambda ) \,+\, B|^{2} \]
where $\lambda_{\mbox{\tiny{SM}}}^{~}$ is the trilinear Higgs self-coupling in
the SM. From the above expression we infer that the cross-section for 
$\gamma\gamma \rightarrow {\rm HH}$ is a quadratic function of
$\lambda \equiv \lambda_{\mbox{\tiny{SM}}}^{~} + \delta\lambda$:
\[ \mbox{\large{$\sigma$}}(\lambda) \,=\, \alpha\lambda^{2} + \beta\lambda + 
\gamma~~~~~~~~~~~~~~~\alpha > 0,~~\gamma > 0 \]

There are various ways to define the sensitivity of the trilinear Higgs
self-coupling. For instance, we can expand around $\mbox{\large{$\sigma$}} =
\mbox{\large{$\sigma$}}_{\mbox{\tiny{SM}}}^{~}$, and express the number of
events as
\[ N \,=\, L\,\mbox{\large{$\sigma$}}_{\mbox{\tiny{SM}}}^{~} \,+\, L\,\delta
\lambda\!\left (\frac{\raisebox{-.4ex}{${\rm d}\mbox{\large{$\sigma$}}$}}
{{\rm d}\lambda}\right )_{\!\lambda\,=\,\lambda_{\mbox{\tiny{SM}}}^{~}} \,+\,
\cdots \]
where $L$ is the integrated luminosity. The sensitivity of $\lambda$ is given
by
\[ \sqrt{N} \,=\, \left |L\,\delta\lambda\!\left (\frac{\raisebox{-.4ex}
{${\rm d}\mbox{\large{$\sigma$}}$}}{{\rm d}\lambda}\right )_
{\!\lambda\,=\,\lambda_{\mbox{\tiny{SM}}}^{~}}\right | \]
i.e.,
\[ \delta\lambda \,=\, \frac{\sqrt{L\,\mbox{\large{$\sigma$}}_{\mbox{\tiny
{SM}}}^{~}}}{L\,({\rm d}\mbox{\large{$\sigma$}}/{\rm d}\lambda )_{\lambda\,=\,
\lambda_{\mbox{\tiny{SM}}}^{~}}^{~}} \,=\, \frac{\sqrt{\mbox{\large{$\sigma$}}_
{\mbox{\tiny{SM}}}^{~}/L}}{({\rm d}\mbox{\large{$\sigma$}}/{\rm d}\lambda )_
{\lambda\,=\,\lambda_{\mbox{\tiny{SM}}}^{~}}^{~}} \] 

\begin{figure}[t]
\begin{center}
\epsfig{file=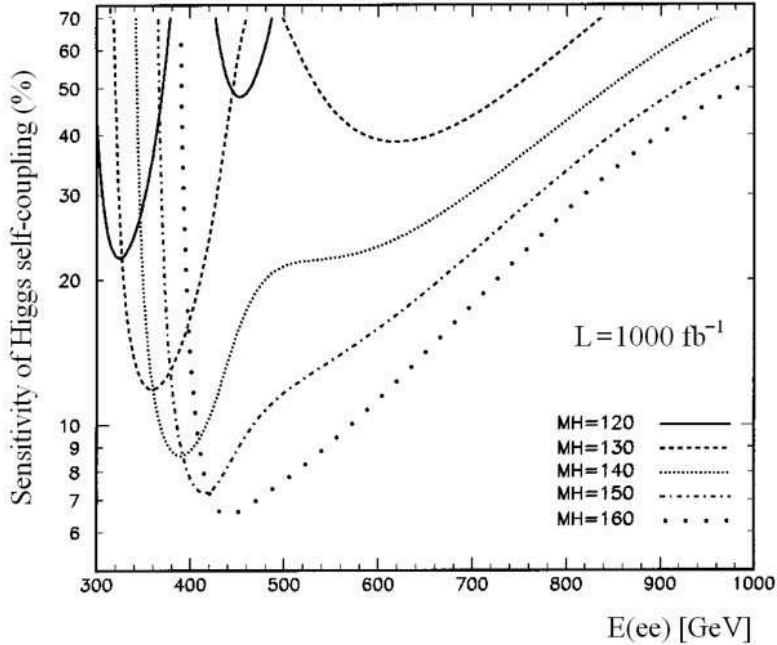,height=0.35\textheight}
\end{center}
\vskip -6mm
\caption{Statistical sensitivity of the trilinear Higgs self-coupling for
various Higgs-boson masses assuming $\int\!{\cal L} = 1000\,fb^{-1}$. Based on
the calculation by R. Belusevic \& G. Jikia described in \cite{belusev1}.}
\label{fig:sensitivity}
\end{figure}

A plot of the trilinear Higgs self-coupling sensitivity in $\gamma\gamma$
collisions, based on the above expression for $\delta\lambda$, is shown in
Fig.~\ref{fig:sensitivity}; for $e^{+}e^{-}$ annihilations, see Fig.~3.8 in
\cite{heinemeyer}. An obvious drawback of the above definition of 
$\delta\lambda$ is that its value becomes unphysically large when the 
derivative ${\rm d}\mbox{\large{$\sigma$}}/{\rm d}\lambda \rightarrow 0$, which
means that one should take into account also the $\lambda^{2}$ term. 

Since the cross-section
$\mbox{\large{$\sigma$}}_{\gamma\gamma\,\rightarrow\,\mbox{\tiny{HH}}}$ does
not exceed 0.4 fb, it is essential to attain the highest possible luminosity,
rather than energy, in order to measure the trilinear Higgs self-coupling.
As shown in \cite{belusev1}, 
appropriate angular and invariant-mass cuts and a $b$-tagging
requirement, which result in a Higgs-pair reconstruction efficiency of about
50\%, would suppress the dominant W-pair and four-quark backgrounds well
below the HH signal. For such a reconstruction efficiency, a center-of-mass
energy E$_{ee}\approx 300$~GeV and $\mbox{\large{$m$}}_{\mbox{\tiny{H}}} = 120$
GeV an integrated $\gamma\gamma$
luminosity $L_{\gamma\gamma} \approx 450$~fb$^{-1}$ would be needed to exclude
a zero trilinear Higgs-boson self-coupling at the 5$\sigma$ level (statistical
uncertainty only). An even higher luminosity is required for an accurate
measurement of this coupling.

\vspace*{0.3cm}
\section{~Higgs couplings to SM particles}
\vspace*{0.3cm}

~~~~In the {\em unitary gauge}, the kinetic term in Eq. (1) can be expressed as
\begin{equation}
(D_{\mu}\Phi )^{\dag}(D^{\mu}\Phi ) \,=\, \frac{\raisebox{-.3ex}{\mbox{\small
{1}}}}{\raisebox{.2ex}{\mbox{\small{2}}}}\mbox{\large{$($}}\partial_{\mu\,}
{\rm H}\mbox{\large{$)$}}^{2} \,+\, \frac{g^{2}}{\raisebox{.2ex}{\mbox{\small
{4}}}}\,(v + {\rm H})^{2}\!\left (W^{+}_{\mu\,}W^{-\,\mu} \,+\, \frac{Z_{\mu\,}
Z^{\mu}}{\mbox{\small{2}}\cos^{2\!}\theta_{\mbox{\tiny{W}}}}\right)
\end{equation}
where $D_{\mu}\Phi$ is the {\em covariant derivative} of $\Phi$ and 
\begin{equation}
\cos\theta_{\mbox{\tiny{W}}} = \frac{g}{\sqrt{g^{2} + g'^{\,2}}}~~~~~~~~~~~~~~~
g\sin\theta_{\mbox{\tiny{W}}} = g'\cos\theta_{\mbox{\tiny{W}}} = {\rm e} 
\end{equation}
($g$ and $g'$ are the electroweak couplings and e is the electric charge). A
comparison with the usual mass terms for the charged and neutral vector bosons
reveals that
\begin{eqnarray}
\mbox{\large{$m$}}_{\mbox{\tiny{W}}}^{~}\!\!&=&\!\!\frac{gv}{\raisebox{.2ex}
{\mbox{\small{2}}}} \\*[2mm]
\mbox{\large{$m$}}_{\mbox{\tiny{Z}}}^{~}\!\!&=&\!\!\frac{gv}{\mbox{\small{2}}
\cos\theta_{\mbox{\tiny{W}}}} \,=\, \frac{\mbox{\large{$m$}}_{\mbox{\tiny{W}}}
^{~}}{\cos\theta_{\mbox{\tiny{W}}}}
\end{eqnarray}
From Eq. (15) we also infer that the {\small\bf Higgs-gauge boson couplings} are
\begin{equation}
\lambda_{\mbox{\tiny{HWW}}} \,\equiv\, \frac{g^{2}v}{\raisebox{.2ex}{\mbox
{\small{2}}}} \,=\, \frac{\mbox{\small{2}}\mbox{\large{$m$}}_{\mbox{\tiny{W}}}
^{\,2}}{\raisebox{.5ex}{$v$}}
\end{equation}
and
\begin{equation}
\lambda_{\mbox{\tiny{HZZ}}} \,\equiv\, \frac{g^{2}v}{\mbox{\small{4}}\cos^{2\!}
\theta_{\mbox{\tiny{W}}}} \,=\, \frac{\mbox{\large{$m$}}_{\mbox{\tiny{Z}}}
^{\,2}}{\raisebox{.5ex}{$v$}}
\end{equation}

Therefore, the Higgs couplings to gauge bosons are proportional to their masses.
This can be readily verified by measuring the production cross-sections in the
Higgs-strahlung and W-fusion processes. At center-of-mass energies 
$s \gg \mbox{\large{$m$}}_{\mbox{\tiny{H}}}^{2}$,
\begin{equation}
   \begin{array}{l}
\mbox{\large{$\sigma$}}(e^{+}e^{-} \rightarrow {\rm HZ}) \,\propto\, \lambda
_{\mbox{\tiny{HZZ}}}^{2}/s \\*[4mm]
\mbox{\large{$\sigma$}}(e^{+}e^{-} \rightarrow {\rm H}\nu\bar{\nu}) \,\propto\,
\lambda_{\mbox{\tiny{HWW}}}^{2}\log\!\mbox{\Large{$($}}s/\mbox{\large{$m$}}_
{\mbox{\tiny{H}}}^{\,2}\mbox{\Large{$)$}}
   \end{array}
\end{equation}
The cross-section $\mbox{\large{$\sigma$}}(e^{+}e^{-} \rightarrow {\rm HZ})
\rightarrow {\rm H}\ell^{+}\ell^{-})$ can be measured independently of the
Higgs-boson decay modes by analyzing the invariant mass of the system
recoiling against the Z boson (see Section 4).

The vector bosons are coupled to the ground-state Higgs field by means of the
covariant derivative (see Eq. (15)). The {\small\bf Higgs-fermion couplings} are
introduced in an {\em ad hoc} way through the {\em Yukawa Lagrangian}
\begin{equation}
{\cal L} \,=\, -g_{\mbox{\tiny{$f$}}}^{~}\overline{\psi}_{\mbox{\tiny{$f$}}}^{~}
\psi_{\mbox{\tiny{$f$}}}^{~}\Phi
\end{equation}
Replacing the Higgs field by its ground-state value, $\Phi \rightarrow v/\sqrt
{2}$ (see Eq. (2)), yields the mass term $-\mbox{\large{$m$}}_{\mbox{\tiny
{$f$}}}^{~}\overline{\psi}_{\mbox{\tiny{$f$}}}^{~}\psi_{\mbox{\tiny{$f$}}}
^{~}$, where $\mbox{\large{$m$}}_{\mbox{\tiny{$f$}}}^{~} = g_{\mbox{\tiny
{$f$}}}^{~}v/\sqrt{2}$. The interaction term in the Lagrangian is obtained by
the replacement $\Phi \rightarrow {\rm H}/\sqrt{2}$:
\begin{equation}
{\cal L}_{\rm int} \,=\, -\,\frac{m_{\mbox{\tiny{$f$}}}}{\raisebox{.5ex}{$v$}}
\,{\rm H}\,\overline{\psi}_{\mbox{\tiny{$f$}}}^{~}\psi_{\mbox{\tiny{$f$}}}^{~}
\end{equation}
We see that, in the Standard Model, all the quarks and charged leptons receive
their masses through {\em Yukawa interactions} with the Higgs field.
Note also that the coupling strength between the Higgs field and the fermion 
\mbox{\small{$f$}} is proportional to the mass of the particle.

Using expression (17), as well as
\begin{equation}
\mbox{\large{$m$}}_{\mbox{\tiny{$f$}}}^{~} = \frac{g_{\mbox{\tiny{$f$}}}^{~}v}
{\sqrt{2}}~~~~~~~~~~{\rm and}~~~~~~~~~~\mbox{\large{$m$}}_{\mbox{\tiny{H}}}^{~}
 = \sqrt{2\lambda}~v
\end{equation}
(see Eq. (4)), we obtain\,\footnote{~We can relate $v$ to the Fermi constant
${\rm G}_{\mbox{\tiny{F}}} = 1.16639\times 10^{-5}~{\rm GeV}^{-2}$ as follows:
\[
\frac{{\rm G}_{\mbox{\tiny{F}}}}{\sqrt{\mbox{\small{2}}}} = \frac{g^{2}}{\mbox
{\small{8}}\mbox{\large{$m$}}_{\mbox{\tiny{W}}}^{\,2}} = \frac{\raisebox{-.3ex}
{\mbox{\small{1}}}}{\mbox{\small{2}}v^{2}}~~~\Rightarrow~~~v =\mbox{\large
{$($}}\sqrt{2}\,{\rm G}_{\mbox{\tiny{F}}}\mbox{\large{$)$}}^{-1/2} \approx
246~{\rm GeV} \] }
\begin{equation}
v \,=\, \frac{\mbox{\large{$m$}}_{\mbox{\tiny{W}}}^{~}}{\raisebox{.4ex}{$g$}/
\mbox{\small{2}}}
 \,=\, \frac{\mbox{\large{$m$}}_{\mbox{\tiny{H}}}^{~}}{\sqrt{2\lambda}} 
 \,=\, \frac{\mbox{\large{$m$}}_{\mbox{\tiny{$f$}}}^{~}}{\raisebox{.4ex} 
{$g_{\mbox{\tiny{$f$}}}^{~}$}/\sqrt{\mbox{\small{2}}}}
\end{equation}
This result is illustrated in Fig.\,2 for $\mbox{\large{$m$}}_{\mbox{\tiny{H}}}
^{~} = 120$ GeV.

The Higgs-fermion couplings can be extracted by measuring the
{\small\bf branching fractions} of the Higgs boson. There are two methods to
determine the Higgs branching fractions: (1) Measure the event rate in the
Higgs-strahlung process for a given final-state configuration and then divide
by the total cross-section; (2) Select a sample of unbiased events in the
Higgs-strahlung recoil-mass peak and determine the fraction of events that
correspond to a particular decay channel. See \cite{heinemeyer} and references
therein for an estimate of the accuracy that can be achieved in such
measurements. As mentioned in Section 3, the Higgs-top coupling can be measured
in the process $e^{+}e^{-} \rightarrow t\bar{t}$ at the pair-production
threshold \cite{fujii}. 

For $\mbox{\large{$m$}}_{\mbox{\tiny{H}}}^{~} \gsim 2\mbox{\large{$m$}}_{\mbox
{\tiny{W}}}^{~}$, the {\small\bf total decay width} of the Higgs boson, $\Gamma
_{\mbox{\tiny{H}}}$, is large enough to be determined directly from the
reconstructed Higgs-boson mass spectrum. The result of such an analysis is
shown in \cite{heinemeyer}. For smaller Higgs-boson masses, $\Gamma_{\mbox
{\tiny{H}}}$ can be determined indirectly by employing the relation between
the total and partial decay widths for a given final state:
\begin{equation}
\Gamma_{\mbox{\tiny{H}}} \,=\, \frac{\Gamma ({\rm H} \rightarrow X)}
{{\rm BR}({\rm H} \rightarrow X)}
\end{equation}
For instance, consider the decay ${\rm H} \rightarrow {\rm WW}^{*}$. One can
directly measure the branching fraction ${\rm BR}({\rm H} \rightarrow
{\rm WW}^{*})$, determine the coupling HZZ in the process $e^{+}e^{-}
\rightarrow {\rm HZ}$, relate the HZZ and HWW couplings based on Eqs. 
(19)--(20), and then use the fact that $\Gamma ({\rm H} \rightarrow {\rm WW})
\propto \lambda_{\mbox{\tiny{HWW}}}^{2}$ to obtain the partial width $\Gamma
({\rm H} \rightarrow {\rm WW}^{*})$ from the information on the HWW coupling.
An accuracy between 4\% and 15\% can be achieved in the determination of
$\Gamma_{\mbox{\tiny{H}}}$ for $\mbox{\large{$m$}}_{\mbox{\tiny{H}}}^{~}$ up to
160 GeV \cite{heinemeyer}.

The decay modes ${\rm H} \rightarrow \bar{b}b$,\,WW can also be measured in 
photon-photon collisions with a precision similar to that expected  from 
analyses based on $e^{+}e^{-}$ data (see, e.g., \cite{asner}). Recall from 
Section 3 that the most accurate way to determine the {\small\bf two-photon
width} $\Gamma ({\rm H} \rightarrow \gamma\gamma )$, which is sensitive to the
Higgs-top coupling, is to combine data from $\gamma\gamma$ and $e^{+}e^{-}$
collisions.

\vspace*{0.3cm}
\section{~The proposed facility}
\vspace*{0.3cm}

~~~~The rich set of final states in $e^{+}e^{-}$ and $\gamma\gamma$ collisions
at a future linear collider (LC) would play an essential role in measuring the 
mass, spin, parity, two-photon width and trilinear self-coupling of the Higgs
boson, as well as its couplings to fermions and gauge bosons; these quantities
are difficult to determine with only one initial state. Furthermore, all the 
measurements made at LEP and SLC could be repeated using highly polarized 
electron beams and at much higher luminosities. For some processes within and 
beyond the Standard Model (e.g., the single and double Higgs-boson production),
the required center-of-mass (CM) energy is considerably lower at the facility
described here than at an $e^{+}e^{-}$ or proton collider. 

The proposed facility could be constructed in two stages, each with distinct 
physics objectives that require particular center-of-mass energies
(see the preprint in \cite{belusev1}): 
\[
       \begin{array}{ll}
{\small\bf Stage~1}:~~~
e^{+}e^{-} \rightarrow {\rm Z,\,WW,\,HZ}~~~~~&~~~~~{\rm E}_{ee} \sim 90~
{\rm to}~250~{\rm GeV} \\*[4mm]
{\small\bf Stage~2}:~~~
\gamma\gamma \rightarrow {\rm H,\,HH}~~~~~&~~~~~{\rm E}_{ee} \sim 160~
{\rm to}~320~{\rm GeV} 
       \end{array} \]
If the $e^{+}e^{-}$ CM energy is increased to about 340 GeV, the top-quark mass
and the Higgs-top coupling could be measured in the process $e^{+}e^{-} 
\rightarrow t\bar{t}$; one expects $\delta\mbox{\large{$m$}}_{t}^{~} \approx
100~{\rm MeV} \approx  0.1\delta\mbox{\large{$m$}}_{t}^{~}({\rm LHC})$.

A schematic layout of an $e^{+}e^{-}$ linear collider is shown in 
Fig.\,\ref{fig:LC} for an X-band machine and in Fig.\,\ref{fig:TESLA} for an
L-band machine. Electron and positron beams are accelerated by a pair of linear
accelerators (linacs) before colliding at an interaction point surrounded by a
detector. The beams are then disposed of, and this machine cycle is repeated at
a rate of 120 Hz (X-band) or 5 Hz (L-band). 

The tunnels containing the main linacs should be sufficiently long to avoid
relocation of the injector complexes each time an energy upgrade takes place.
The production and testing of the accelerating structures and rf sources needed
for such an upgrade, and the subsequent installation of the latter, can be
carried out with minimal disruption to the data-taking process if the 
klystrons, modulators and pulse compressors are placed in a separate tunnel
(see Fig.\,\ref{fig:tunnels}). 

It is also envisaged that `bypass lines' for low-energy beams\,\footnote{~
Bypass lines are not needed at an L-band accelerator facility.} would be
employed to accumulate data at the Z resonance in the process $e^{+}e^{-}
\rightarrow{\rm Z}$. This data could be used to regularly calibrate the
detector and measure the luminosity of the accelerator. Assuming a geometric
luminosity $L_{e^{+}e^{-}}^{~} \approx 5\times 10^{33}$ cm$^{-2}$\,s$^{-1}$ at
the Z resonance, approximately $2\times 10^{9}$ Z bosons could be produced in 
an operational year of $10^{7}$ s; this is about 200 times the entire LEP 
statistics.

\vspace*{0.3cm}
\section{~Photon collider}
\vspace*{0.3cm}

~~~~The idea of using counter-directed electron linacs to create a gamma-gamma
collider can be traced back to an article by P. Csonka published in 1967 
\cite{csonka}. The seminal work on photon colliders by I. Ginzburg et al.
\cite{ginzburg} describes in detail a method for obtaining $\gamma\gamma$ and
$e\gamma$ collisions by Compton backscattering of laser light on high-energy
electrons. 

\begin{figure}[!h]
\begin{center}
\epsfig{file=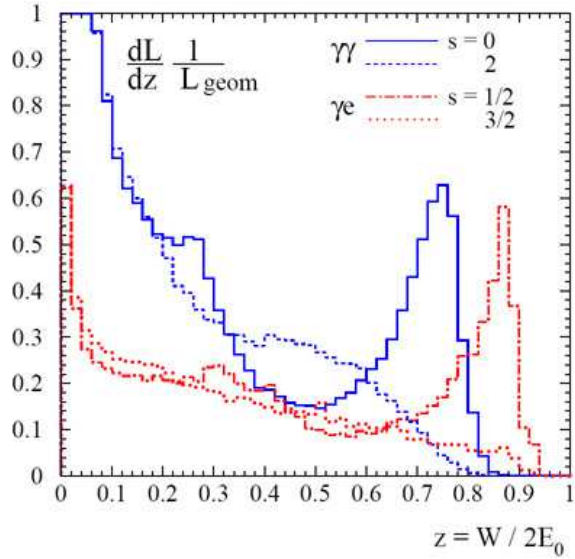,height=0.3\textheight}
\end{center}
\vskip -7mm
\caption{Simulated $\gamma\gamma$ and $e\gamma$ luminosity spectra
\cite{telnovACTA} (see Section 14).}
\label{fig:spectra}
\end{figure}

The backscattered photons have energies comparable to those of the incident
electrons (see Fig.\,\ref{fig:spectra}), and follow their direction with some
small angular spread of the order of $1/\gamma$, where $\gamma$
is the Lorentz factor. The spatial spread of the photons is
approximately $d/\gamma$ at a distance $d$ from the Compton interaction point
(CIP). Both the energy spectrum and polarization of the backscattered photons
depend strongly on the polarizations of the incident electrons and laser
photons. The key advantage of using $e^{-}e^{-}$ beams at a $\gamma\gamma$
collider is that they can be polarized to a high degree.

At CIP, the electron beam is about 10 times wider than it would be at the $ee$
collision point in the absence of a laser beam. However, since the
backscattered photons follow the direction of the incident electrons, they are
automatically `focused' to their collision point.

The absence of beam-beam effects in $\gamma\gamma$ collisions means that it
is not necessary to have very flat linac beams. The spectral luminosity of
$\gamma\gamma$ collisions strongly depends on beam characteristics, but only
through the parameter $\rho$, the ratio of the intrinsic transverse spread of
the photon beam to that of the original electron beam: $\rho \equiv d/\gamma
\sigma_{e}^{~}$. In this expression, $d$ is the distance between CIP and the
$\gamma\gamma$ collision point, and $\sigma_{e}^{~}$ is the radius that a round
Gaussian linac beam would have at the collision point in the absence of a laser
beam. As $\rho$ increases, the monochromaticity of the luminosity distribution
improves (because the lowest-energy photons, which scatter at the largest 
angles, do not pass through the collision point), but the total luminosity 
decreases. For a typical photon collider, the optimal value of $d$ is a few 
millimeters \cite{TESLA}.
 
Assuming that the mean number of Compton interactions of an electron in a laser
pulse (the Compton conversion probability) is 1, the {\em conversion
coefficient} $k \equiv n_{\gamma}^{~}/n \approx 1 - \mbox{\small{e}}^{-1} =
0.63$, where $n_{e}$ is the number of electrons in a 'bunch' and $n_{\gamma}
^{~}$ is the number of scattered photons. The luminosity of a gamma-gamma
collider is then
\begin{equation}
{\cal L}_{\gamma\gamma} \,=\, (n_{\gamma}^{~}/n_{e}^{~})^{2\,}{\cal L}_{ee}
 \,\approx\, (0.63)^{2\,}{\cal L}_{ee}
\end{equation}
where ${\cal L}_{ee}$ is the {\small\bf geometric luminosity} of electron 
beams:
\begin{equation}
{\cal L}_{ee} \,=\,\frac{\gamma n_{e}^{\,2}{\rm N}_{b}f}{4\pi\sqrt{\mbox{\large
{$\varepsilon$}}_{x}\beta_{x}\mbox{\large{$\varepsilon$}}_{y}\beta_{y}}}
\end{equation}
In this expression, $\mbox{\large{$\varepsilon$}}_{x},\mbox{\large
{$\varepsilon$}}_{y}$ are the {\em beam emittances}, $\beta_{x},\beta_{y}$ are
the horizontal and vertical {\em beta functions}, respectively, N$_{b}$ is the 
number of bunches per train, and $f$ is the beam collision frequency. In
the high-energy part of the photon spectrum, ${\cal L}_{\gamma\gamma}\sim 0.1
{\cal L}_{ee}$. However, if beams with smallest possible emittances and 
stronger beam focusing in the horizontal plane are used, then ${\cal L}_{\gamma
\gamma}$ could, in principle, be made higher than ${\cal L}_{e^{+}e^{-}}$
\cite{telnovACTA} (see also Section 14).

\vspace*{0.3cm}
\section{~X-band accelerator complex}
\vspace*{0.3cm}

\begin{figure}[h]
\begin{center}
\epsfig{file=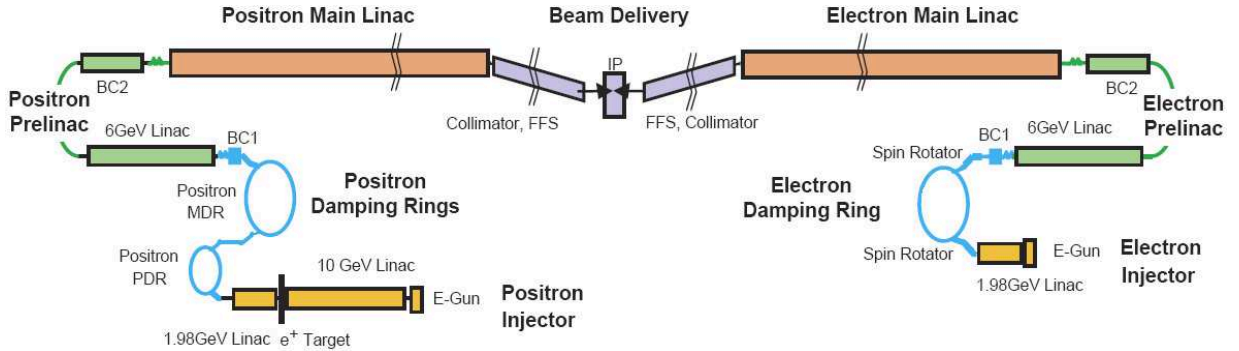,height=0.187\textheight}
\end{center}
\vskip -7mm
\caption{Schematic layout of an X-band linear $e^{+}e^{-}$ collider \cite{GLC}.
With a crossing angle at the interaction point (IP), separate beam lines can be
used to bring the disrupted beams to their respective dumps, thereby enabling
post-IP diagnostics.}
\label{fig:LC}
\end{figure}

~~~~A schematic layout of an X-band linear $e^{+}e^{-}$ collider is shown in 
Fig.\,\ref{fig:LC}. The current X-band (11.4 GHz) rf technology has been
developed mainly at KEK and SLAC \cite{adolphsen}. The choice of this
technology is motivated by the cost benefits of having relatively low rf energy
per pulse and high accelerating gradients. The ongoing effort to develop
high-gradient X-band structures is essential for the eventual construction of a
CLIC-type linear accelerator \cite{ellis}

\begin{figure}[t]
\begin{center}
\epsfig{file=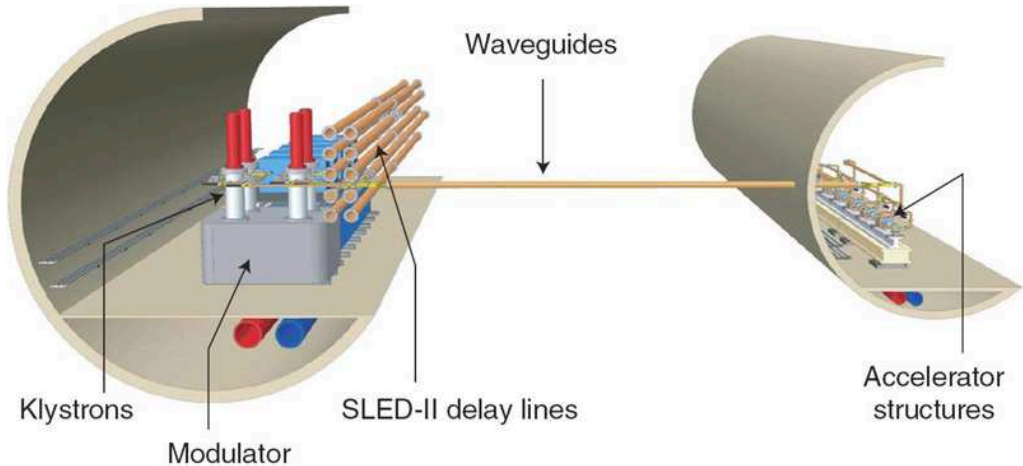,height=0.25\textheight}
\end{center}
\vskip -4mm
\caption{Dual tunnels for an X-band linear collider \cite{GLC}.}
\label{fig:tunnels}
\end{figure}

The tunnels containing the rf sources and accelerating structures are sketched
in Fig.\,\ref{fig:tunnels}. A single rf unit contains a modulator that drives a
pair of 60 MW klystrons, each of which generates 1.6 $\mu$s rf pulses at 120
Hz. An rf compression system enhances the peak power of the klystrons by a
factor of about three, and produces 400 ns pulses that match the accelerator
structure requirements.\,\footnote{~It takes $\sim$120 ns to fill each rf
cavity with an accelerating field.  The remaining period of $\sim$280 ns is
used to accelerate a `train' of electron bunches, which has a total length of
about 270 ns.} The resulting 380 MW, 400 ns pulses feed six 0.6 m long
accelerator structures, producing a 68 (54) MV/m unloaded (loaded) gradient in
each structure.

To reduce power consumption, it is proposed to use klystrons with
superconducting solenoidal focusing \cite{ogitsu}. The XL4 klystron
developed at SLAC, for instance, could initially be adapted for this purpose
\cite{sprehn}.\,\footnote{~To avoid using power-consuming solenoid
electromagnets, SLAC and KEK have developed klystrons with periodic permanent
magnet (PPM) focusing \cite{sprehn}. All PPM klystrons built so
far suffer chronic rf breakdown in the output section, which manifests itself
by a loss of transmitted power that develops over several hundred ns
\cite{adolphsen}. An alternative approach to the development of high-power
klystrons is therefore proposed in the main text.}

A comprehensive review of the status of X-band accelerator technology is given
in \cite{adolphsen}. Significant advances have been made in pulsed HV and rf
power generation, high gradient acceleration and wakefield supression. The rf
sources and accelerating structures of the facility proposed here would have
properties very similar to those described in \cite{adolphsen}. The ultimate
design of rf cavities, however, will depend on the outcome of the ongoing
effort to develop 100 MeV/m X-band structures for a CLIC-type linear collider.

\vspace*{0.3cm}
\section{~Laser system for an X-band machine}
\vspace*{0.3cm}
 
~~~~In order to attain maximum luminosity, every electron bunch in the 
accelerator should collide with a laser pulse of sufficient intensity for
$63\%$ of the electrons to undergo a Compton scattering. This requires a laser
system with high average power, capable of producing pulses that would match
the temporal spacing of electron bunches. The laser power is minimized when 
the Rayleigh range of the laser focus and the laser pulse width are matched to
the electron bunch length. The proposed collider would have about 95 100-micron
bunches separated by 2.8 ns, with 120 trains per second. This means that 
$95\times 120 = 11,400$ laser pulses with a duration of approximately 1 ps must
be produced every second. To avoid nonlinear electrodynamic effects, the
maximum pulse energy should not exceed 1 joule. Therefore, the laser system
ought to deliver about 11 kW of average power in pulses of a terawatt peak 
power, matched to the linac bunch structure. 

\begin{figure}[t]
\begin{center}
\epsfig{file=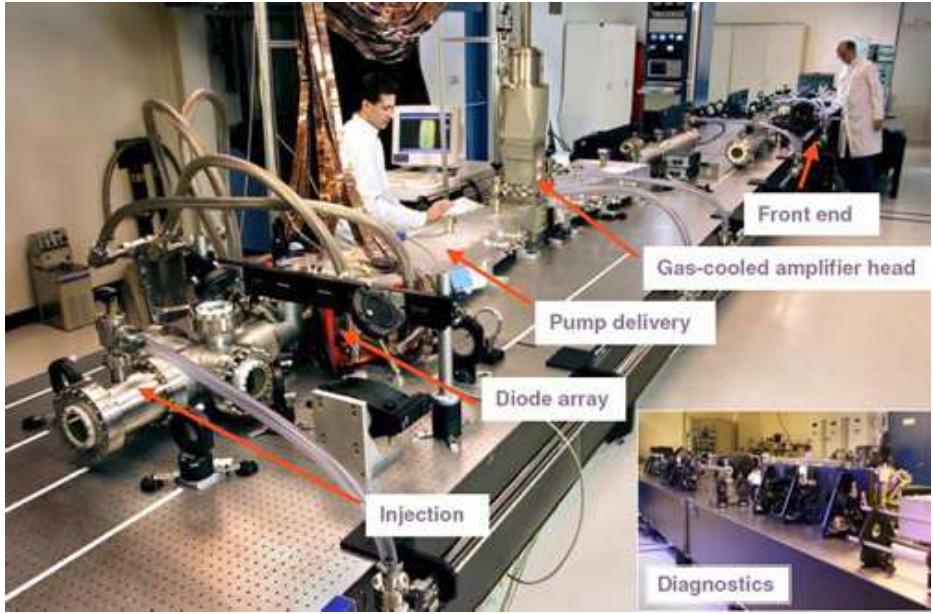,height=0.33\textheight}
\end{center}
\vskip -4mm
\caption{Photograph of the {\em Mercury laser}. Credit: C. Bibeau and LST.} 
\label{fig:Mercury}
\end{figure}

These requirements could be satisfied, for instance, by modifying the 
{\small\bf Mercury laser} developed at the Lawrence Livermore Lab (LLNL) for a 
laser fusion application, as proposed by J. Gronberg and his collaborators
\cite{gronberg, gronberg1} (see Fig.\,\ref{fig:Mercury}).
The {\em Mercury laser} is designed to generate 100-J pulses
with a width of 2$-$10 ns at a rate of 10 Hz. An initial low-energy
picosecond pulse is `chirped' to produce a nanosecond pulse. Once amplified in
the {\em Mercury}, the pulse can be recombined back to the picosecond level as
long as it has retained a Gaussian profile after amplification. The technique
of chirped-pulse amplification, invented by D. Strickland and G. Mourou in
1985, is at the heart of all laser designs that can produce terawatt peak 
power. The {\em Mercury laser} would use Ytterbium-doped crystals (Yb:S-FAP),
pumped by diode arrays, to deliver 100~J at 10 Hz with 10\% efficiency for
converting wall plug power to laser light. The long upper state lifetime of the
crystal ($\tau = 1.26$ ms) allows the laser to be pumped more slowly, reducing
the required peak diode power.

An array of 12 {\em Mercury lasers} can be fired sequentially to match the 120
Hz repetition rate of the collider. The single 100-Joule pulse can then be 
subdivided and time delayed using a series of optical splitters and delay lines
to produce a train of 95 1-Joule pulses separated by 2.8 ns. This would provide
one laser pulse for every electron bunch in the linac. A system of focusing
mirrors in the interaction region allows each laser pulse to be used twice, 
thus providing laser pulses for each arm of the linear accelerator.

Apart from the issue of long-term laser stability, this approach is not optimal
because only one out of twelve lasers is used at a given moment, which means
that a dozen independent diode arrays are required. It would be more practical
if a number of special lasers with high repetition rate were used in parallel.
Alternatively, one could employ an optical {\em free electron laser} or doped 
{\em fiber amplifiers} (see Summary).

\vspace*{0.3cm}
\section{~Interaction region and beam dump}
\vspace*{0.3cm}

\begin{figure}[!t]
\begin{center}
\epsfig{file=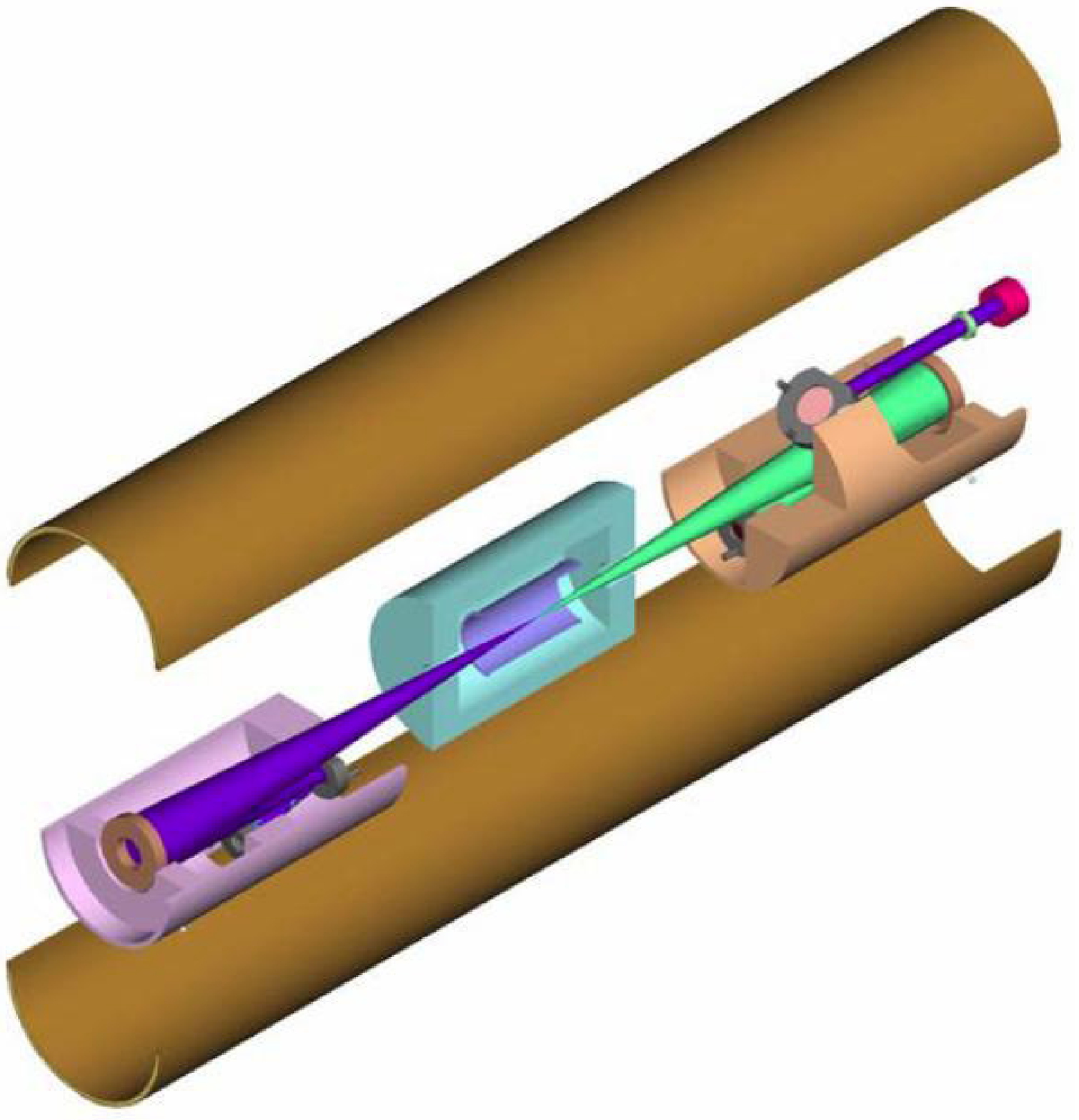,height=0.33\textheight}
\epsfig{file=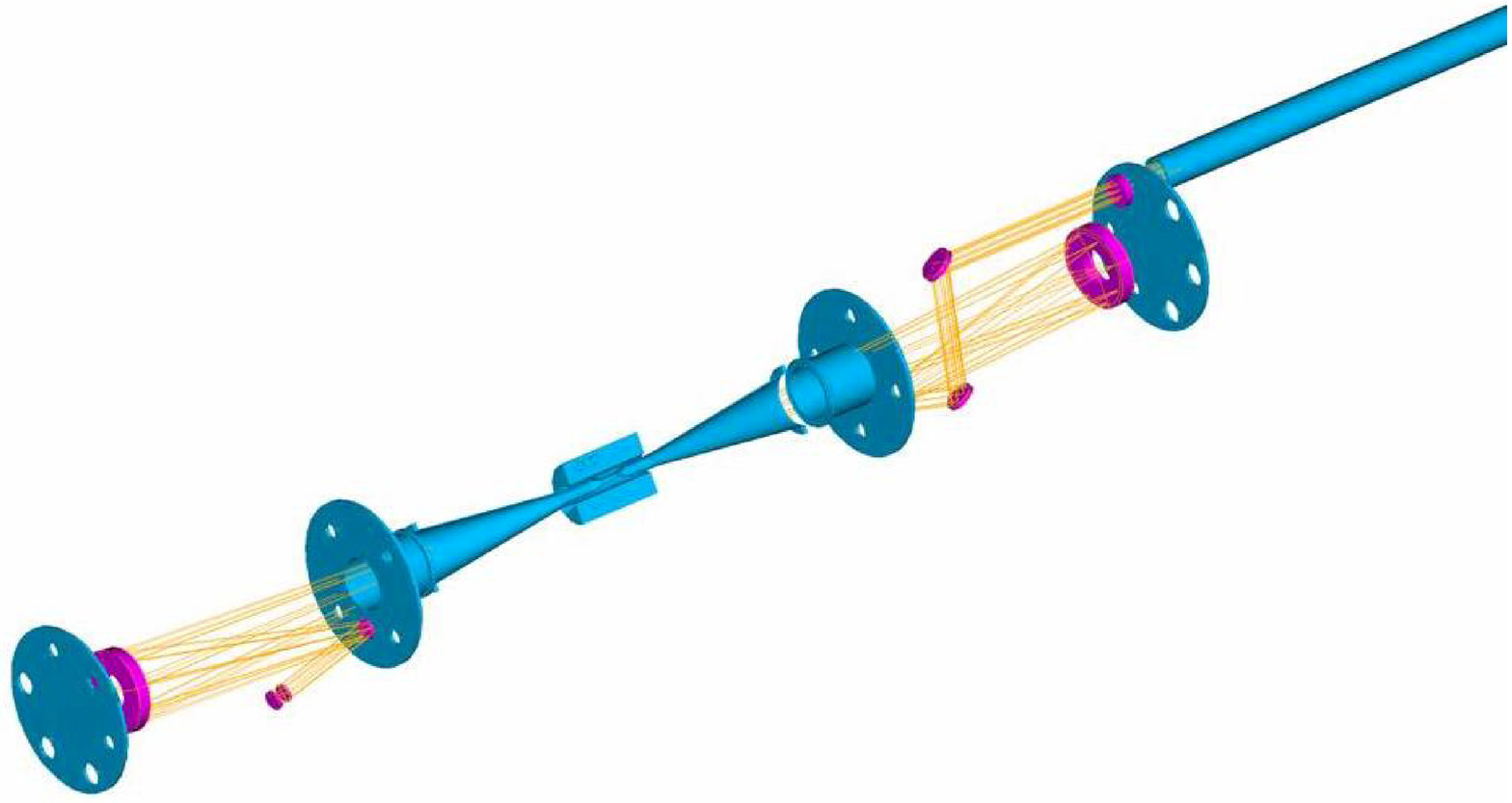,height=0.24\textheight}
\end{center} 
\vskip -4mm
\caption{Optics assembly at the $\gamma\gamma$ interaction region. Electron
beams and most of the background particles pass through the central hole in
each of the two end-mirrors. Elements of the assembly were designed, simulated 
and prototyped at LLNL \cite{skulina, gronberg1}.}
\label{fig:optics}
\end{figure}

~~~~The {\small\bf optics assembly} for the interaction region at a photon
collider shown in Fig.\,\ref{fig:optics} satisfies the following requirements:
(a) the laser beam must be nearly co-linear with the electron beam;
(b) the latter must pass through the final focusing optics;
(c) the beams of electrons and laser photons must simultaneously be at the
Compton interaction point;
(d) the duration of the laser pulse must correspond to the electron bunch 
length. 

The Compton scattering of laser photons on high-energy electrons results in a
large energy spread in the electron beam. At the interaction point (IP), this 
leads to a large angular spread of the outgoing beam due to the beam-beam 
interaction. For nominal beam and laser parameters, the extraction beam pipe
must therefore have an aperture of about $\pm 10$ milliradians.

To remove the disrupted beams, one can use the {\small\bf crab-crossing
scheme} proposed by R. Palmer. In this scheme, the beams are collided at a
{\em crossing angle} of about 10 to 20 milliradians. The same luminosity as in
head-on collisions can be obtained by tilting the electron bunches with respect
to the direction of the beam motion. 

The aperture of the extraction beam pipe and the physical size of the final
focusing magnet set a lower limit on the crossing angle of the colliding beams.
The minimum crossing angle is about 25 milliriadians if a final focusing
quadrupole magnet with a compensating coil to minimize the fringe field is used
\cite{parker}.\,\footnote{~The fringe field from the final focusing magnet must
be minimized to prevent low-energy particles, which are swept away by the
field, from causing radiation- and heat-related problems.} This is somewhat 
larger than the crossing angles envisaged for the proposed ILC and NLC 
$e^{+}e^{-}$ colliders.

The `feedback' system for bringing the beams into collision relies on post-IP 
{\small\bf beam position monitors} (BPMs) that measure the beam-beam deflection
at the collision point. Because of the energy spread in a highly disrupted 
beam, conventional BPMs may not provide sufficient resolution due to electric
noise. Moreover, it is not possible to steer such a beam without large beam
losses. This implies that the extraction line at a $\gamma\gamma$ collider will
be a straight vacuum pipe, which precludes some post-IP diagnostics such as
precise measurement of the beam energy and polarization. 

Much of the extracted-beam power will be in the form of high-energy photons 
that have a very narrow angular spread. This would result in a large amount of
energy being deposited within a small volume of the water {\small\bf beam 
dump}, causing vaporization of H$_{2}$O. A possible solution to this problem
would be to convert the photons to $e^{+}e^{-}$ pairs in a gas target situated
before the dump. In order to decrease the flux of backward-scattered neutrons,
a volume filled with hydrogen or helium gas could be placed just before the gas
target (see \cite{shekhtman} and \cite{telnovACTA} for more detail).

Huge savings in construction cost could be achieved if the crossing angle and
the beam dump are exactly the same for the operation of the accelerator in the
$e^{+}e^{-}$ and $\gamma\gamma$ collision modes. The beam dump described in
\cite{shekhtman} is designed with this in mind. The part of the extraction line
containing a chicane --- which provides vertical displacement and dispersion 
needed for continuous measurements of the beam energy spectrum and polarization
at an $e^{+}e^{-}$ collider --- could be replaced with a straight vacuum pipe 
for the operation in the $\gamma\gamma$ mode. 

\vspace*{0.3cm}
\section{~L-band (TESLA-type) accelerator complex}
\vspace*{0.3cm}

\begin{figure}[h]
\begin{center}
\epsfig{file=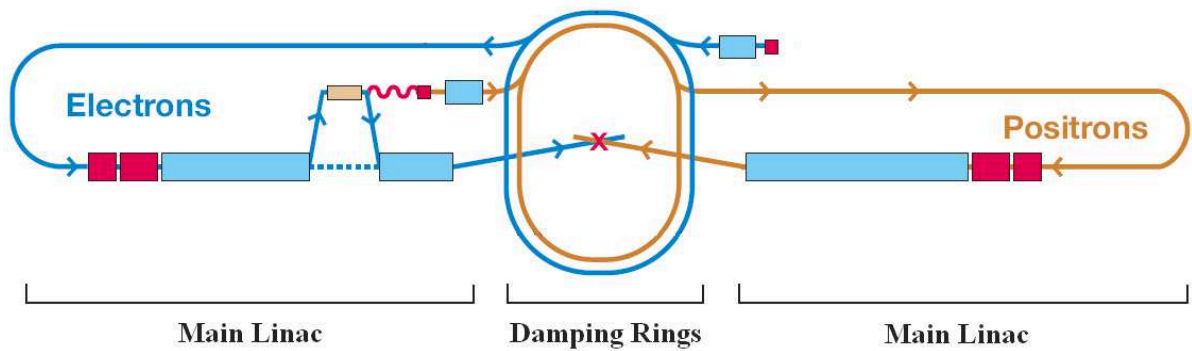,height=0.192\textheight}
\end{center}
\vskip -4mm
\caption{Schematic layout of an L-band (TESLA-type) linear collider; credit:
ILC collaboration. Each linac consists of a few thousand superconducting
cavities placed within cryogenic vessels to form cryomodules (see
Fig.\,\ref{fig:cryomodule}). Liquid helium is used to cool cavities to
$-271^{\circ}$C.}
\label{fig:TESLA}
\end{figure}

\begin{figure}[!t]
\begin{center}
\epsfig{file=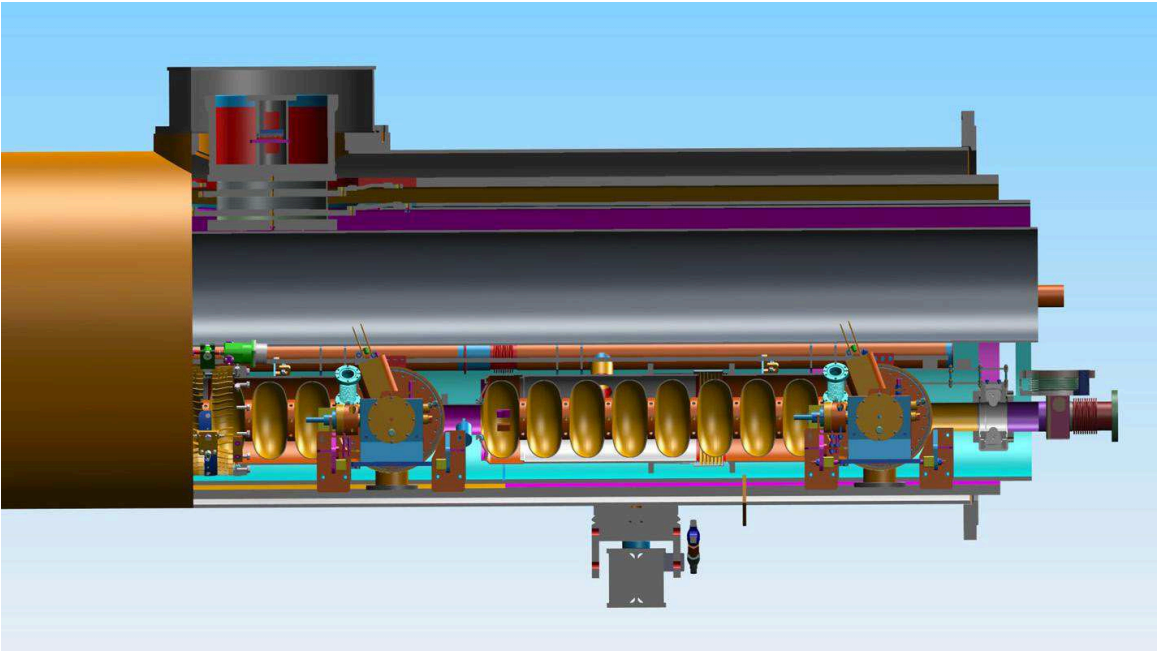,height=0.35\textheight}
\end{center}
\vskip -4mm
\caption{An inside view of a cryomodule for the ILC. Credit: Fermilab.}
\label{fig:cryomodule}
\end{figure}

~~~~A detailed description of the current design for the {\em International
Linear Collider} (ILC) can be found in \cite{ILC} (see Fig.\,\ref{fig:TESLA}).
This design, based on the superconducting technology originally developed at
DESY, uses L-band (1.3 GHz) rf cavities that have average accelerating
gradients of 31.5 MeV/m. A single superconducting niobium cavity is about 1 m
long. Nine cavities are mounted together in a string and assembled into a
common low-temperature cryostat or {\em cryomodule} (see 
Fig.\,\ref{fig:cryomodule}), which is 12.652 m long.\,\footnote{~A 20-GeV linac
containing 116 cryomodules is currently under construction at DESY. This
facility will serve both as an X-ray free electron laser (XFEL) and an
important testbed for the ILC.}

The ILC main linacs are composed of rf units, each of which is formed by three 
contiguous cryomodules containing 26 nine-cell cavities. Every unit has an rf
source, which includes a pulse modulator, a 10 MW multi-beam klystron, and a
waveguide system that distributes the power to the cavities. 

A TESLA-type design offers some advantages over the X-band technology:

\vspace*{1mm}
\noindent
$\bullet$~~Wakefields are drastically reduced due to the large size of the rf
cavities, which means that cavity alignment tolerances can be relaxed;\\
\noindent 
$\bullet$~~Superconducting rf cavities can be loaded using a long rf pulse
(1.5 ms) from a source with low peak rf power;\\ 
\noindent
$\bullet$~~`Wall-plug to beam' power transfer efficiency is about twice that
of X-band cavities;\\ 
\noindent
$\bullet$~~The long rf pulse allows a long bunch train ($\sim 1$ ms), with many
bunches ($\sim 2600$) and a relatively large bunch spacing ($\sim 370$ ns). A
trajectory correction (feedback) system within the train can therefore be used
to bring the beams into collision.
 
However, in contrast to a compact, high-gradient X-band machine, a collider 
based on the current TESLA-type design would be characterized by low 
accelerating gradients ($\lsim 30$ MeV/m), damping rings that a few kilometers
in circumference, and a technologically challenging cryogenic system (see
Fig.\,\ref{fig:cryomodule}) that requires a number of surface cryogenic plants.

If the initial operation of the proposed facility is in the $\gamma\gamma$
mode, there would be no need for an $e^{+}$ source. Two electron damping 
rings could then be built inside a single tunnel. For operation at the nominal
$e^{+}e^{-}$ luminosity, a positron damping ring would later replace one of the
electron rings. In any case, the present design of the positron source
(see Fig.\,\ref{fig:TESLA}) should be greatly simplified.

\vspace*{0.3cm}
\section{~Laser system for an L-band machine}
\vspace*{0.3cm}

~~~~To use {\em Mercury lasers} with a TESLA-type machine, which has a bunch
structure very different from that of an X-band collider,  would require
prohibitively large average power and development of special electro-optical 
switches. Alternatively, one could take advantage of the large inter-bunch 
spacing in the TESLA-type linac and use a $\sim$100-m long optical 
cavity that would allow a single laser pulse to travel around a ring, colliding
with all the bunches. However, it is extremely difficult to achieve required
precision in the construction of such a cavity. A particularly undesirable
feature of this scheme is a large `dead angle' ($\pm 95$ mrad) for
particle tracking inside the detector \cite{telnovACTA}.

A possible source of primary photons for a $\gamma\gamma$ collider is an
optical {\em free electron laser} (FEL). The radiation produced by an FEL has
a variable wavelength, and is fully polarized either circularly or linearly
depending on whether the undulator is helical or planar, respectively. The 
required time structure of laser pulses can be achieved by using a modified 
design of the main linac as a driving accelerator for the FEL. A free electron
laser for the photon collider at a TESLA-type machine is described in 
\cite{saldin}.

\vspace*{0.3cm}
\section{~Luminosity and backgrounds at a $\gamma\gamma$ collider}
\vspace*{0.3cm}

~~~~Since the cross-sections 
$\mbox{\large{$\sigma$}}_{\gamma\gamma\,\rightarrow\,\mbox{\tiny{HH}}}$ and
$\mbox{\large{$\sigma$}}_{e^{+}e^{-}\,\rightarrow\,\mbox{\tiny{HHZ}}}$ do not 
exceed 0.4 fb, it is essential to attain the highest possible luminosity,
rather than energy, in order to measure the trilinear Higgs self-coupling. If
beams with smallest possible emittances and stronger beam focusing in the
horizontal plane are used, then the $\gamma\gamma$ luminosity ${\cal L}_{\gamma
\gamma}$ could, in principle, be made higher than ${\cal L}_{e^{+}e^{-}}$, as
explained in what follows \cite{telnovACTA}.

Assuming the `nominal' ILC beam parameters for the operation of the accelerator
in the $e^{+}e^{-}$ mode \cite{ILC},\,\footnote{~The `nominal' ILC beam 
parameters are: ${\rm N}_{b} = 2\times 10^{10}$, $\sigma_{z} = 0.3$\,mm,
$f = 14100$\,Hz, $\varepsilon_{x} = 10^{-5}$\,m and $\varepsilon_{y} = 4\times
10^{-8}$\,m. While it is feasible to obtain $\beta_{y} \sim\sigma_{z}=0.3$\,mm,
the minimum value of $\beta_{x}$ is restricted, by chromo-geometric aberrations
in the final-focus system and for the above horizontal emittance, to about
5\,mm.} the expected $\gamma\gamma$ luminosity in the high-energy part of the
photon spectrum
\[ {\cal L}_{\gamma\gamma}(\mbox{\small{high-energy peak}}) \,\sim\, 3.5\times 
10^{33}\,{\rm cm}^{-2}\,{\rm s}^{-1} \,\sim\, 0.2{\cal L}_{e^{+}e^{-}} \]
where ${\cal L}_{e^{+}e^{-}} = 2\times 10^{34}\,{\rm cm}^{-2}\,{\rm s}^{-1}$ is
limited by collision effects (beamstrahlung and beam instabilities). At a 
photon collider with center-of-mass energies $\lsim 500$ GeV, and for electron
beams that are not too short, coherent pair production is suppressed due to the
broadening and displacement of the electron beams during their collision
\cite{TESLA}. In this case, ${\cal L}_{\gamma\gamma}$ is limited {\em only} by
the transverse area of the beam (note that its vertical size is much smaller 
than the horizontal):
\begin{equation}
{\cal L}_{\gamma\gamma} \,\propto\, (\mbox{\large{$\sigma$}}_{x}^{~}\mbox
{\large{$\sigma$}}_{y}^{~})^{-1}~~~~~~~~~~~~~~~\mbox{\large{$\sigma$}}_{x,y}
^{~} \,=\, \sqrt{\beta_{x,y}(\mbox{\large{$\varepsilon$}}_{x,y}/\gamma )}
\end{equation}
as can be seen from expressions (27) and (28) in Section 8. 

\begin{figure}[!t]
\begin{center}
\epsfig{file=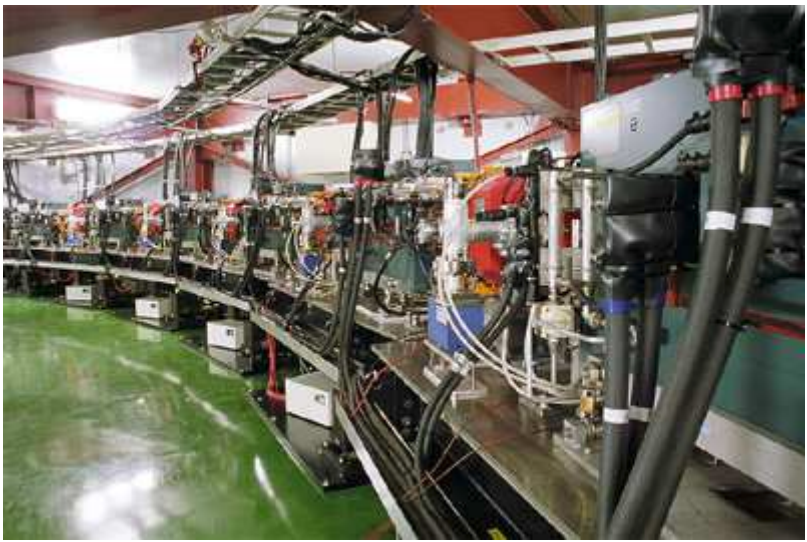,height=0.30\textheight}
\end{center}
\vskip -4mm
\caption{Arc section of the ATF Damping Ring, which produces the world's
smallest-emittance beams. The layout of the magnets in the arc sections is
designed to achieve small equilibrium emittances. The wiggler magnets in the
straight sections of the ring shorten the damping time. Fast kicker magnets and
DC septum magnets are used for beam extraction. Credit: KEK.}
\label{fig:ATF}
\end{figure}

The beam emittances in Eq. (29) are determined by various physics effects 
inside a damping ring (see Fig.\,\ref{fig:ATF}). If the synchrotron radiation
is dominated by the ring's wiggler parameters (large $F_{\rm w}$), and if the
quantum excitation by the wiggler is not too large compared with that in the 
arcs, then from Eqs. (33) and (14) in \cite{emma} it follows that the 
horizontal beam emittance $\mbox{\large{$\varepsilon$}}_{x}$ could be 
significantly reduced by using a wiggler with short period and a judicially 
chosen value of the peak field (in order to preserve the damping time). The
vertical emittance $\mbox{\large{$\varepsilon$}}_{y}$ is {\em not} determined
by the wiggler, but by optics errors that are not easily characterized. 
Assuming that both $\mbox{\large{$\varepsilon$}}_{x}$ and $\mbox{\large
{$\varepsilon$}}_{y}$ could be reduced by about a factor of five compared with
their 'nominal' ILC values, ${\cal L}_{\gamma\gamma}$ would then exceed
${\cal L}_{e^{+}e^{-}}$:
\[ {\cal L}_{\gamma\gamma}(\mbox{\small{high-energy peak}}) \,\sim\, 2.5\times
10^{34}\,{\rm cm}^{-2}\,{\rm s}^{-1} \,\sim\, 1.2{\cal L}_{e^{+}e^{-}} \]
To obtain this result it was also assumed that $\beta_{x} = 1.7$\,mm, and that
the distance between the interaction and conversion regions is 1\,mm. Simulated
luminosity spectra for these parameters are shown in Fig.\,\ref{fig:spectra}.

At a $\gamma\gamma$ collider, the spectrum of photons after Compton scattering
is broad, with a characteristic peak at maximum energies (see 
Fig.\,\ref{fig:spectra}). The low-energy part of the spectrum is produced by 
multiple Compton scattering of electrons on photons inside laser beams. 

The Compton-scattered photons can have circular or linear polarizations,
depending on their energies and the polarizations of the initial electrons and
laser light. For instance, the scattered photons have an average helicity
$\langle\lambda_{\gamma}\rangle \neq 0$ if either the laser light has a 
circular polarization ${\rm P}_{\!c} \neq 0$ or the incident electrons have a
mean helicity $\langle\lambda_{e}\rangle \neq 0$. In the case $2{\rm P}_{\!c}
\lambda_{e}= -1$, which results in a good monochromaticity of the backscattered
photon beam, the average degree of circular polarization of the photons within
the high-energy peak of the luminosity distribution is over 90\%.

Since the polarization of Compton-scattered photons depends strongly on their
energy, the {\small\bf luminosity spectrum} has to be measured separately for 
different polarization states. When both photons are {\em circularly 
polarized}, the process $\gamma\gamma\rightarrow e^{+}e^{-},\,\mu^{+}\mu^{-}$ 
is particularly well suited for measuring the spectral luminosity \cite{pak}. 
This process has a cross-section of a few pb for a total $\gamma\gamma$ angular
momentum $|J_{z}| = 2$. A precision of about 0.1\% is expected in one year of
running, which is better than the accuracy needed for the Higgs-boson studies 
described in this note. To measure the luminosity spectrum in the $|J_{z}| = 0$
configuration, the helicity of one of the photon beams can be inverted by
simultaneously changing the signs of the helicities of both the laser and 
electron beams. For the product of photon {\em linear polarizations}, the
spectral luminosity can be measured in the above process by studying the 
azimuthal variation of the cross-section at large angles \cite{pak, TESLA}.

Undisrupted electron beams at a $\gamma\gamma$ collider can be steered using a 
fast feedback system that measures their deflection (see Section 11). Once the 
electron beams are brought into collision, the laser will be turned on. The
scattered photons follow the direction of the incident electrons.

Multiple Compton scattering of electrons on photons leads to a low-energy
`tail' in the energy spectrum of the electrons. At the interaction point, this
results in a large deflection angle of the $e^{-}e^{-}$ beams. Due to a finite
crossing angle (see Section 11), the outgoing beams are also deflected 
vertically by the solenoidal magnetic field of the detector. Fig.\, 19 in
\cite{bechtel} shows the angular spread of an outgoing electron beam right
after the interaction point and at $z = 2.8$ m. The problem of 'stabilizing'
beam-beam collisions, and hence the $\gamma\gamma$ luminosity, is discussed in 
\cite{telnovACTA}. 

The {\small\bf backgrounds} at a photon collider caused by beam-beam effects 
in the interaction region have been simulated considering both the 
{\em incoherent} particle-particle and {\em coherent} particle-beam
electromagnetic (QED) interactions described in \cite{TESLA}. Another
significant source of background is due to backscattering of particles. The
hadronic structure of the photon arises from the possibility that it can either
split into a quark-antiquark pair or transform into a vector meson, with the
probability of about 1/200. At the expected ILC $\gamma\gamma$ luminosity, for
instance, the average number of hadronic background events per one bunch 
collision is about two \cite{TESLA}. The above backgrounds influence data
acquisition and analysis, as well as the operation of various detector
components, as discussed in \cite{TESLA, bechtel}.

\vspace*{0.3cm}
\section{~Summary and Acknowledgements}
\vspace*{0.3cm}

~~~~The rich set of final states in $e^{+}e^{-}$ and $\gamma\gamma$ collisions
at a future linear collider would play an essential role in measuring the mass,
spin, parity, two-photon width and trilinear self-coupling of the Higgs boson,
as well as its couplings to fermions and gauge bosons (see Sections 3 to 6);
these quantities are difficult to determine with only one initial state. For
some processes within and beyond the Standard Model, the required 
center-of-mass energy is considerably lower at the facility described here than
at an $e^{+}e^{-}$ or proton collider.

Since the cross-sections
$\mbox{\large{$\sigma$}}_{\gamma\gamma\,\rightarrow\,\mbox{\tiny{HH}}}$ and
$\mbox{\large{$\sigma$}}_{e^{+}e^{-}\,\rightarrow\,\mbox{\tiny{HHZ}}}$ do not
exceed 0.4 fb, it is essential to attain the highest possible luminosity,
rather than energy, in order to measure the trilinear Higgs self-coupling. If
beams with smallest possible emittances and stronger beam focusing in the
horizontal plane are used, then the luminosity ${\cal L}_{\gamma\gamma}$ could
be made higher than ${\cal L}_{e^{+}e^{-}}$ (see Section 14).

\begin{figure}[!h]
\vskip -37mm
\setlength{\unitlength}{1cm}
\begin{picture}(10,10)
\put(0.25,0){\epsfig{file=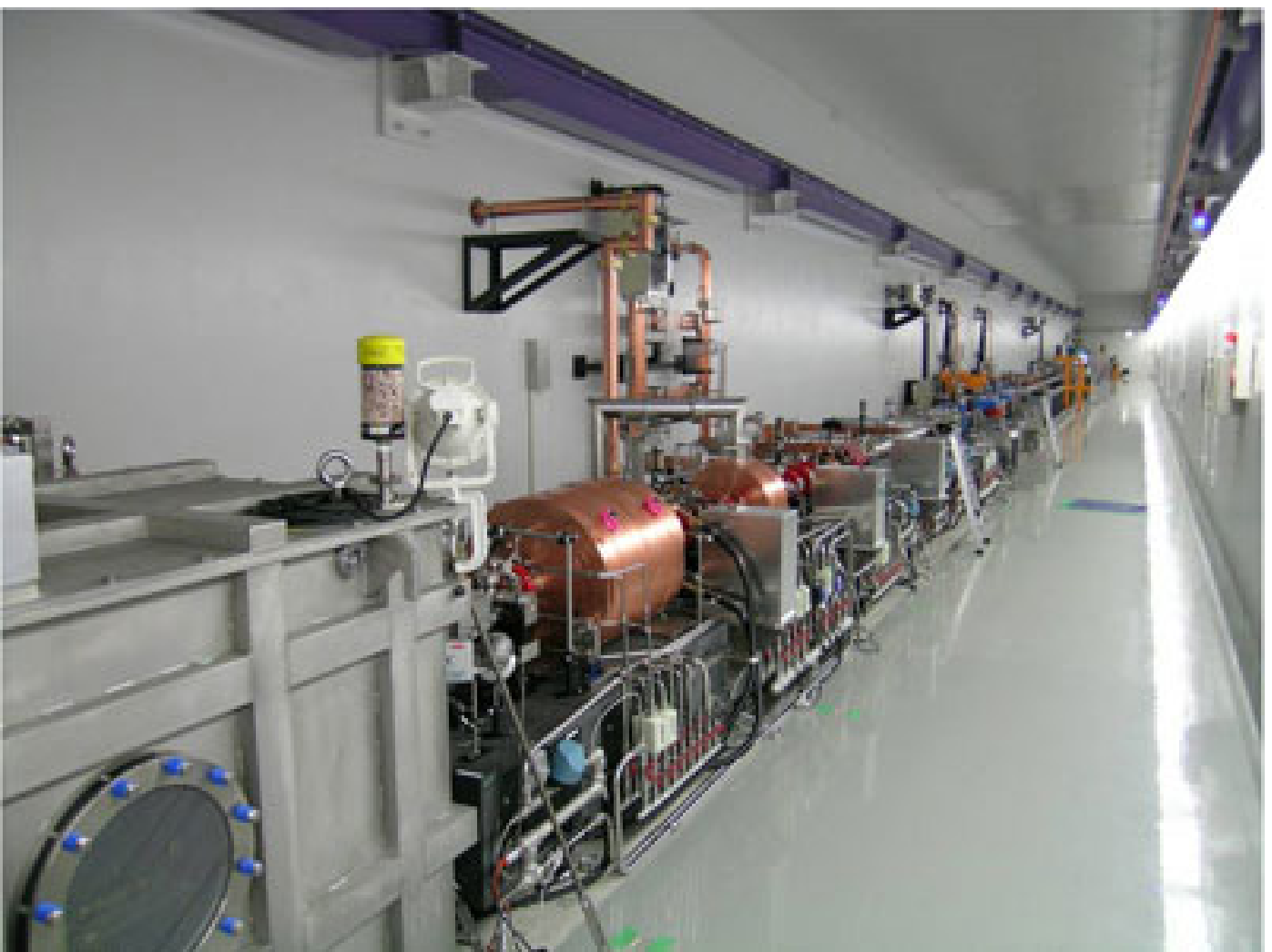,width=0.48\textwidth}}
\put(8.5,0){\epsfig{file=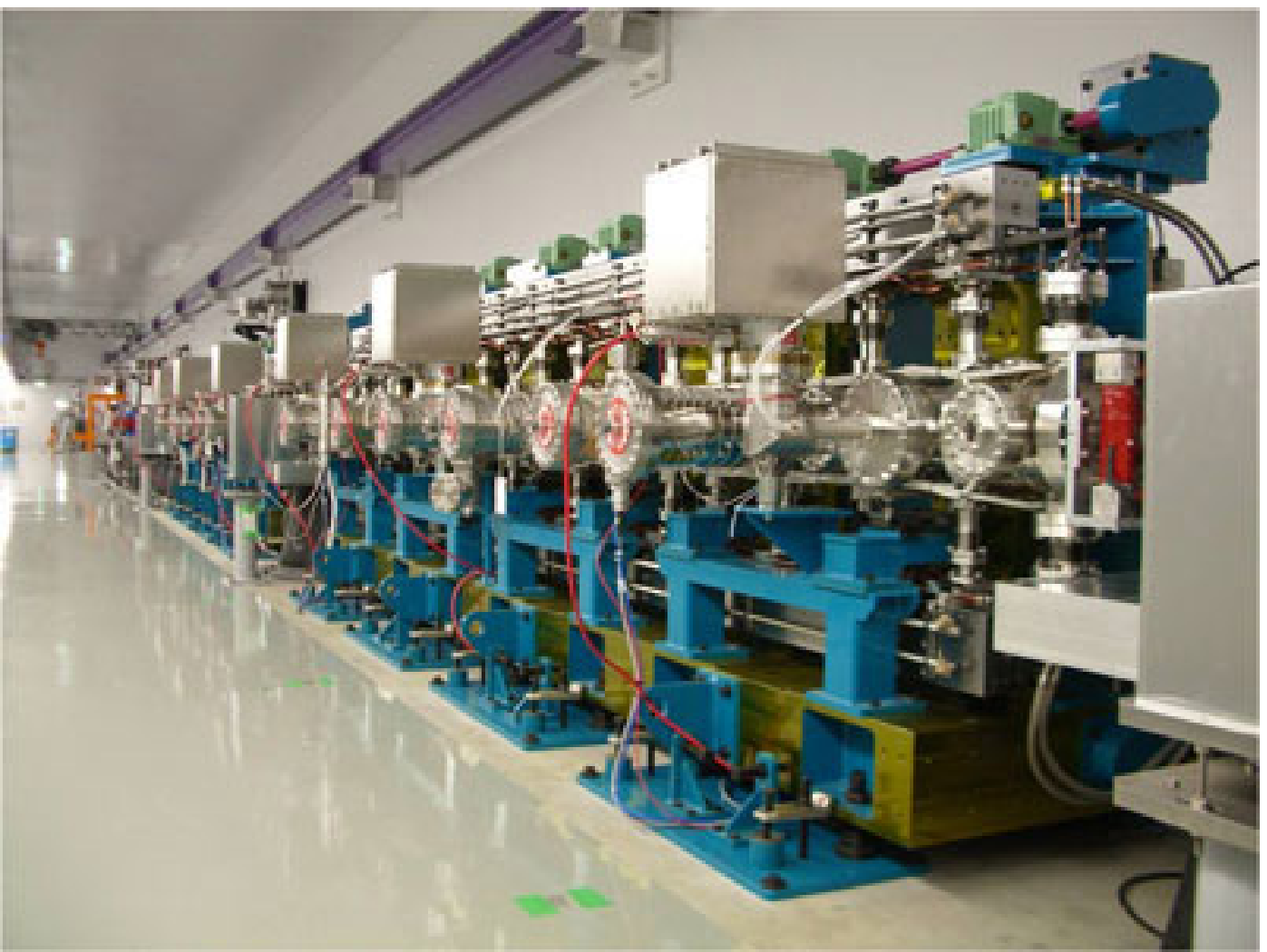,width=0.48\textwidth}}
\end{picture}
\caption{The 8-GeV electron linac (left) and the in-vacuum undulator of the
X-ray free electron laser (XFEL) at the SPring-8 facility in Japan
\cite{shintake}. The C-band rf system of the linac can produce an accelerating
gradient of 32 MV/m using 50 MW klystrons.}
\label{fig:XFEL}
\end{figure} 

The proposed $e^{+}e^{-}/\gamma\gamma$ collider would be constructed in three 
stages, each with a distinct physics objective that requires a particular 
center-of-mass energy (see Section 7 and the preprint in \cite{belusev1}).
Together with LHC, such a facility would bridge the gap between the present 
high-energy frontier and that accessible to a TeV-scale $e^{+}e^{-}$ or muon
collider. 

An L-band (TESLA-type) linear collider offers some advantages over an X-band
machine (see Section 12). However, in contrast to a compact, high-gradient 
X-band accelerator, a collider based on the current TESLA-type design would be
characterized by low accelerating gradients ($\lsim 30$ MeV/m), damping rings
that are a few kilometers in circumference, and a technologically challenging
cryogenic system that requires a number of surface cryogenic plants. One should
also bear in mind that the ongoing effort to develop 100 MeV/m X-band 
accelerating structures is essential for the eventual construction of a 
TeV-scale linear collider based on the CLIC design.

If the initial operation of the proposed facility is in the $\gamma\gamma$
mode, there would be no need for an $e^{+}$ source. Two electron damping
rings could then be built inside a single tunnel in the case of a TESLA-type
machine. For operation at the nominal $e^{+}e^{-}$ luminosity, a positron 
damping ring would later replace one of the electron rings.

A possible source of primary photons for a $\gamma\gamma$ collider is: 
an optical {\em free electron laser} (FEL), diode-pumped {\em solid state 
lasers}, or doped {\em fiber amplifiers} (see Sections 10 and 13). The 
radiation produced by an FEL has a variable wavelength, and is fully polarized
either circularly or linearly. Each of the X-ray free electron lasers (XFELs)
currently under development at SLAC (S-band), DESY (L-band) and the SPring-8 
facility (C-band; see Fig.\,\ref{fig:XFEL}) can serve as a testbed for both an
optical FEL and the main linac of a future linear collider.\,\footnote{~The
wavelength $\lambda$ of FEL radiation is determined by $\lambda \approx \lambda
_{u}/2\gamma^{2}$, where $\gamma$ is the Lorentz factor of the electron beam
and $\lambda_{u}$ is the periodic length of the undulator. Although an optical
FEL requires a much smaller electron linac and a considerably simpler undulator
than an XFEL, the charge per electron beam bunch has to be sufficiently large
($\sim 4$ nC) to produce photon pulses of $\sim 1$ J. Suitable high-intensity 
and low-emittance rf guns have already been developed \cite{michelato}.}
An optical FEL could be placed in a separate tunnel connected to the 
experimental hall housing the detector.

Elements of the optics assembly for the interaction region at a photon collider
were designed, simulated and prototyped at LLNL (see Section 11).
The Compton scattering of laser photons on high-energy electrons results in a
large energy spread in the electron beam. At the interaction point, this
leads to a large angular spread of the outgoing beam due to the beam-beam
interaction. To remove the disrupted beams, one can use the crab-crossing 
scheme described in Section 11. Huge savings in construction cost could be
achieved if the crossing angle and the beam dump are exactly the same for the
operation of the accelerator in the $e^{+}e^{-}$ and $\gamma\gamma$ collision 
modes.

\vspace*{0.8cm}
\begin{center}
{\large\bf Acknowledgements}
\end{center}
\vspace*{0.4cm}

I am grateful to K. Fl\"{o}ttmann, K. Fujii, S. Fukuda, J. Gronberg, 
K. Hagiwara, T. Higo, S. Hiramatsu, G. Jikia, S. Kazakov, S. Matsumoto,
A. Miyamoto, Y. Okada, A. Seryi, D. Sprehn, N. Toge, A. Wolski and K. Yokoya
for many useful discussions regarding various aspects of this proposal.

\end{document}